\documentclass{aa}  

\usepackage{rotating}
\usepackage{graphicx}
\usepackage{txfonts}
\usepackage{multirow}
\usepackage{amsmath}
\usepackage{xcolor}
\usepackage{soul}
\usepackage[breaklinks, colorlinks, citecolor=blue, linkcolor=blue]{hyperref}
\usepackage{soul}
\usepackage{caption}
\usepackage{subcaption}

\newcommand{\cheops}{{CHEOPS}} % SH: I removed the white spaces after the names so facilitate the writing 

\newcommand{\spitzer}{{Spitzer}}
\newcommand{\hale}{{Hale}}
\newcommand{\gaia}{{Gaia DR3}}

\newcommand{\kepler}{{Kepler}}
\newcommand{\kt}{{K2}}
\newcommand{\harps}{{HARPS }}
\newcommand{\system}{{K2-138}}
\newcommand{\rsun}{$R_\odot$}
\newcommand{\msun}{$M_\odot$}
\newcommand{\mearth}{$M_\oplus$}
\newcommand{\rearth}{$R_\oplus$}
\newcommand{\supdagger}{\textsuperscript{\textdagger}}

\newcommand{\lang}[1]{#1}

\begin{document}

\title{New ephemerides and detection of transit-timing variations in the \system\ system using high-precision \cheops{} photometry}
\titlerunning{Ephemerides and TTVs of the \system\ system}
\authorrunning{Vivien et al.}

\author{
    H.~G. Vivien\inst{1} $^{\href{https://orcid.org/0000-0001-7239-6700}{\includegraphics[scale=0.5]{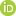}}}$
    \and
    S. Hoyer\inst{1} $^{\href{https://orcid.org/0000-0003-3477-2466}{\includegraphics[scale=0.5]{figures/orcid.jpg}}}$
    \and
    M. Deleuil\inst{1}
    \and
    S. Sulis\inst{1}
    $^{\href{https://orcid.org/0000-0001-8783-526X}{\includegraphics[scale=0.5]{figures/orcid.jpg}}}$
    \and
    A. Santerne\inst{1} $^{\href{https://orcid.org/0000-0002-3586-1316}{\includegraphics[scale=0.5]{figures/orcid.jpg}}}$
    \and
    J.~L. Christiansen\inst{2} $^{\href{https://orcid.org/0000-0002-8035-4778}{\includegraphics[scale=0.5]{figures/orcid.jpg}}}$
    \and\\
    K.~K. Hardegree-Ullman\inst{3} $^{\href{https://orcid.org/0000-0003-3702-0382}{\includegraphics[scale=0.5]{figures/orcid.jpg}}}$
    \and
    T.~A. Lopez
}

\institute{
    Aix Marseille Univ, CNRS, CNES, Institut Origines, LAM, Marseille, France
    \and
    Caltech/IPAC-NASA Exoplanet Science Institute, Pasadena, CA 91125, USA
    \and
    Steward Observatory, The University of Arizona, Tucson, AZ, 85721, USA
}

\date{}

% For abstract:
%   - Context (optional)
%   - Aims
%   - Methods
%   - Results
%   - Conclusion (optional)

\abstract{
Multi-planet systems are a perfect laboratory for constraining planetary formation models. A few of these systems present planets that come very close to mean motion resonance, potentially leading to significant transit-timing variations (TTVs) due to their gravitational interactions. Of these systems, \system{} represents a excellent laboratory for studying the dynamics of its six small planets (with radii ranging between $\sim1.5$ -- $3.3\,$\rearth), as the five innermost planets are in a near 3:2 resonant chain. 
}{
In this work, we aim to constrain the orbital properties of the six planets in the \system{} system by monitoring their transits with CHaracterising ExOPlanets Satellite (\cheops). We also seek to use this new data to lead a TTV study on this system.
}{
We obtained twelve light curves of the system with transits of planets $d$, $e$, $f,$ and $g$. With these data, we were able to update the ephemerides of the transits for these planets and search for timing transit variations.
}{
With our measurements, we reduced the uncertainties in the orbital periods of the studied planets, typically by an order of magnitude. This allowed us to correct for large deviations, on the order of hours, in the transit times predicted by previous studies. This is key to enabling future reliable observations of the planetary transits in the system. We also highlight the presence of potential TTVs ranging from 10 minutes to as many as 60 minutes for planet $d$.
}
{}

\keywords{Planetary systems --- Planets and satellites: detection --- Planets and satellites: fundamental parameters --- Stars: individual: K2-138 --- Techniques: photometry}

\maketitle
%\onecolumn

%--------------------------------------------------------------------
% Section
%--------------------------------------------------------------------

\section{Introduction}

Multi-planetary systems account for about 25\% of confirmed planets to date\footnote{\label{note1} Available systems gathered from \url{exoplanet.eu}}. The exoplanetary system known to host the largest number of planets so far is Kepler-90, with eight planets \citep{Shallue_2018}. The vast majority of these systems are composed of small planets, super-Earths and mini-Neptunes, in the innermost part of the systems ($<100$ days periods). These packed systems are the ideal laboratory for testing and calibrating models of planetary formation: originating from the same original disc, the composition of these planets gives us clues about the region of the disc where they are likely to have formed. The orbital architecture of the system gives us information about the orbital evolution they may have undergone \citep{Mishra_2021, mishra2023, Luque2023}.

In these compact systems, the gravitational interactions between planets can be strong, as predicted by \citet{agol2005}. These interactions result in variations of differing magnitude (from a few seconds to nearly half an hour) in the transit times of the planets. Measuring these transit time variations \citep[TTV,][]{miralda-escude2002, agol2005, holman2005} requires high photometric accuracy and a high measurement cadence in order to accurately determine transit times as well as ingress and egress. Studies of such systems can also benefit from a long time coverage to better identify which planet is transiting at a given time and track the transit time evolution. As TTVs are gravitational effects, they can \lang{be used to measure} the mass of planets from photometric data alone, providing insights into their possible composition.

To date, there are 39 systems known to host five or more planets, but only a handful with a resonant chain \citep{Lissauer_2014}. This phenomenon occurs when multiple successive planets are in (or close to) mean motion resonance \citep[MMR,][]{lissauer2011, fabrycky2014}, namely, the planet period ratios are close to a ratio of small integers $(r+1):r$. Among these systems, we observe chains of at least three planets only in: Kepler-90 \citep{Kepler-90_Cabrera}, HD~219134 \citep{HD219134}, TRAPPIST-1 \citep{TRAPPIST-1_Luger, TRAPPIST-1-refined}, K2-138 \citep{christiansen2018, acuna2022}, HD~110067 \citep{HD110067}, Kepler-80 \citep{Shallue_2018}, TOI-1136 \citep{TOI-1136_Dai}, TOI-178 \citep{TOI-178-Leleu, TOI-178-Delrez}, and HD~158259 \citep{HD158259}. Such observations question formation models that predict a breaking of the chain soon after the dispersion of the protoplanetary disk via dynamical instabilities. The exact nature of these instability processes is however still debated \citep{Goldberg2022}. The study of these systems therefore allows us to probe the architecture of these particular systems at a time when the system is in a stable phase.

\system{} is a prime example of a highly packed resonant system. Originally described in \citet{christiansen2018} using \kt\ data, it was found to host five sub-Neptunes close to 3:2 MMR. A better characterization of the planets' masses was obtained by \citet{lopez2019} using \harps\ observations. Using \spitzer, \lang{\citet[][hereafter HU21]{hardegree-ullman2021}} confirmed an outer sixth planet, not in the resonant chain. Finally, \lang{\citet[][hereafter A22]{acuna2022}} investigated the composition of the planets and found that the inner planets present an increasing water content with distance from the host star, while the outer planets have an approximately constant water content.

The cadence of the \kt{} photometry (8 to 10 min) did not allow for a TTV analysis of the system for which TTV amplitudes between 2.1 and 6.5 min are expected \citep{lopez2019}. We therefore observed the \system{} system with \cheops{} \citep{Benz2021} to obtain high-precision photometry of the target and further explore the presence of TTVs. The higher cadence of \cheops, 60~sec in the case of K2-138 should therefore be sufficient to evaluate TTVs of even two minutes. 

The paper is organized as follows. First, we describe the observations used in this work in Sect.~\ref{sec:observations} and the \cheops{} data modeling in Sect.\ref{sec:results}. We then present improved ephemerides of the system in Sect.~\ref{sec:ephemerides} and our TTV analysis in Sect.~\ref{ssec:ttv_modeling}. Finally, we present our conclusions in Sect.~\ref{sec:conclusion}.

%--------------------------------------------------------------------
% Section
%--------------------------------------------------------------------

\begin{figure*}
    \centering
    \includegraphics[width=1\hsize]{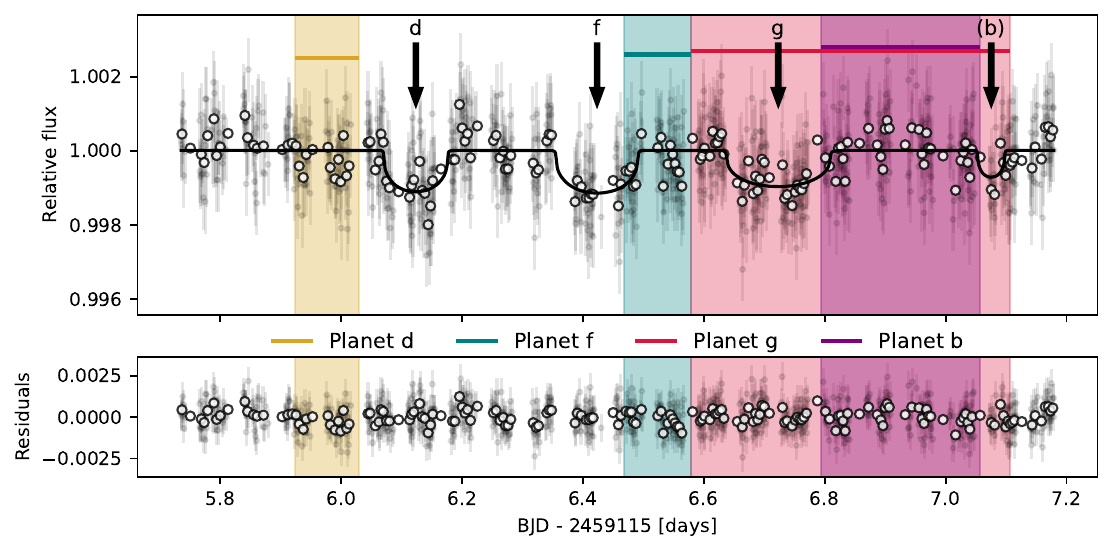}
    \caption{Detrended light curve of \cheops{} visit 501. The colored regions correspond to the predicted transit time (with $\pm1\sigma$ errors) of planets $d$, $f$, $g,$ and $b$ according to the ephemerides from A22. The detrended \cheops{} data are represented by the light gray points, with their five-minute bins shown with the white dots. The black solid curve shows the fit of the four planet transits and arrows show the planets associated to each transit. Planet $b$ is written in parentheses, as the depth of the event ($\sim550$~ppm) doesn't appear compatible with prior data ($\sim250$~ppm).}
    \label{fig:4_transit}
\end{figure*}

\section{Observations and modeling}\label{sec:observations}

\subsection{Stellar and planetary properties}\label{ssec:stellar_prop}

\renewcommand{\arraystretch}{1.2} % Better spacing for readability
\begin{table}[h]
\caption{Properties of the \system{} system used in this work.}
\centering
\begin{tabular}{ l r }
    \hline
    \multirow{2}{1.5 cm}{Alternative names} & \gaia{} 2413596935442139520 \\
     & EPIC 245950175 \\
    \hline\hline
    Parameter & Value \\
    \hline
    Mass [\msun] & $0.891^{+0.017}_{-0.027}$ \\
    Radius [\rsun] & $0.834^{+0.011}_{-0.01}$ \\
    $T_{\mathrm{eff}}$ [K] & $5354.7^{+27.9}_{-21.2}$ \\
    $\log(g)$ [cgs] & $4.55^{+0.02}_{-0.02}$ \\
    $\rho$ [$\rho/\rho_\odot$] & $1.534^{+0.081}_{-0.090}$ \\
    $[\mathrm{Fe/H}]$ [dex] & $0.07 \pm 0.05$ \\
    Age [Gyr] & $3.3^{+2.4}_{-3.2}$ \\
    Distance [pc] & $201.5 \pm 1.9$ \\
    Limb darkening $h_1$ & $0.4906^{+0.0075}_{-0.0071}$ \\
    Limb darkening $h_2$ & $0.2080^{+0.0045}_{-0.0047}$ \\
    %R.A. [J2000] & 23h 15m 47.7686067840 \\
    %DEC. [J2000] & -10 50 58.895461284 \\
    \hline
\end{tabular}
\label{table:stellar_parameters}
\end{table}

A22 conducted a full spectroscopic analysis using stellar modeling to derive the atmospheric and fundamental parameters of the host star. The improved host star's parameters helped determining the possible composition of the planets in the system. To use completely homogeneous priors during our analysis, we used the stellar and planetary parameters derived in that work. For sake of clarity, the stellar parameters are given in Table~\ref{table:stellar_parameters}.  We note that the values reported by A22 for the planets are consistent with those in \citet{lopez2019}. 

The system is thought to be made up of six planets: $b$, $c$, $d$, $e$, $f,$ and $g$. Planets $b$ through $f$ are found to be very close the 3:2 resonance, thus forming one of the longest known chains. The periods of these planets are, respectively, 2.35, 3.56, 4.40, 8.26, 12.76, and 41.97~days.

\subsection{Previous photometric observations}\label{ssec:prior_obs}

Available observations of the system prior to this work include the original \kt{} data from campaign 12. This photometry was acquired over a 79 day period, between December 15, 2016  and  March 4, 2017, using the long cadence mode of the \kepler{} spacecraft \citep{Howell2014}. The \spitzer{} Space Telescope subsequently observed the system using the InfraRed Array Camera (IRAC) for 11 hours between March 15 and 16, 2018  (DDT 13253; PI: J.~L. Christiansen). These observations were targeting planet $g$, extending coverage to the tenth epoch since discovery. Finally, the \hale{} Telescope at the Palomar Observatory observed \system{} with the Wide-field InfraRed Camera (WIRC) on two occasions, August 31, 2019 and November 4, 2019. Both observations were aimed at planet $d$, and observed epochs 182 and 194, respectively \citep[][hereafter B22]{boyle2022}.

\subsection{\cheops{} observations}\label{ssec:cheops_data}

\system\ was observed by \cheops{} between 2020 July 22 and 2020 October 6 over twelve visits with durations between 8h42m and 34h44m as part of program 17 (PI: T.A. Lopez). The original schedule of observations was planned to acquire transits of planets $d$, $e$, $f,$ and $g$. Due to the brightness of the star ($V=12.2$) an exposure time of 60 seconds was used. In total, \cheops{} observations account for slightly more than five and a half days, covering fifteen transits of four planets in the system: four transits for planet $d$, three for planet $e$, six for planet $f$, and one for planet $g$. Details about each visit can be found in Table~\ref{table:cheops_files}. For the sake of simplicity, each light curve is given an identifier based on the last three digits of the corresponding file name.

\renewcommand{\tabcolsep}{3pt} % Making it fit on the page
\begin{table*}
\caption{Detail of the twelve \cheops{} observations.}
\centering

\begin{tabular}{l l r r r r r r}
    \hline
    Planets & ID & \texttt{File name} & \texttt{OBSID} & Start date [UTC] & Duration [h:m:s] & Frames & Efficiency [\%] \\
    \hline\hline
    $d$ & 101 & \texttt{CH\_PR210017\_TG000101\_V0200} & \texttt{1213611} & 2020-08-27T09:09:31 & 8:42:15 & 483 & 92.3 \\
    $d$ ($b$) & 102 & \texttt{CH\_PR210017\_TG000102\_V0200} & \texttt{1229610} & 2020-09-12T14:52:36 & 8:42:15 & 471 & 90.0 \\
    $d$ & 103 & \texttt{CH\_PR210017\_TG000103\_V0200} & \texttt{1253194} & 2020-10-04T02:59:36 & 8:42:15 & 418 & 79.9 \\
    $e$ ($b$, $c$) & 201 & \texttt{CH\_PR210017\_TG000201\_V0200} & \texttt{1202221} & 2020-08-17T07:25:31& 8:48:15 & 449 & 84.8 \\
    $e$ & 202 & \texttt{CH\_PR210017\_TG000202\_V0200} & \texttt{1218498} & 2020-08-25T14:41:30 & 8:59:15 & 487 & 90.1 \\
    $e$ & 203 & \texttt{CH\_PR210017\_TG000203\_V0200} & \texttt{1251541} & 2020-09-27T13:37:36 & 8:59:15 & 491 & 90.9 \\
    $f$ & 301 & \texttt{CH\_PR210017\_TG000301\_V0200} & \texttt{1180664} & 2020-07-22T02:49:09 & 8:54:15 & 272 & 50.8 \\
    $f$ & 302 & \texttt{CH\_PR210017\_TG000302\_V0200} & \texttt{1189091} & 2020-08-03T19:04:30 & 9:20:16 & 341 & 60.8 \\
    $f$ ($d$) & 303 & \texttt{CH\_PR210017\_TG000303\_V0200} & \texttt{1206371} & 2020-08-16T13:07:30 & 9:06:15 & 460 & 84.1 \\
    $f$ ($b$) & 304 & \texttt{CH\_PR210017\_TG000304\_V0200} & \texttt{1218441} & 2020-08-29T08:11:30 & 9:23:16 & 523 & 92.7 \\
    $f$ & 305 & \texttt{CH\_PR210017\_TG000305\_V0200} & \texttt{1253418} & 2020-10-06T15:39:14 & 9:05:15 & 352 & 64.4 \\
    $d$, $f$, $g$ ($b$) & 501 & \texttt{CH\_PR210017\_TG000501\_V0200} & \texttt{1237897} & 2020-09-23T05:29:36 & 34:43:58 & 1911 & 91.7 \\
    \hline
\end{tabular}
\tablefoot{Planets in parenthesis were predicted by the ephemerides \citet{acuna2022} but not detected in \cheops{} observations. The efficiency refers to the fraction of observing time over the full duration of the visit.}
\label{table:cheops_files}

\end{table*}

Light curves were obtained after the calibration and correction of the images by the \cheops{} automatic data reduction pipeline \citep[DRPv13.1,][]{hoyer2020}. This version of the DRP automatically delivers the target's photometry for four different aperture radii: \texttt{DEFAULT}$=$25\,px, \texttt{RINF}$=$22.5\,px, \texttt{RSUP}$=$30\,px, and \texttt{OPTIMAL,} which is computed for each target based the target's flux and contamination of the field of view. For \system, the \texttt{OPTIMAL} aperture was computed to be of 24\,px. To compute the latter, the DRP uses the \emph{GAIA DR2} catalog \citep{gaia_dr2} to simulate the position and flux of putative contaminants in the field of view of the target. For \system, no such contaminant is located within any of the apertures. The closest contaminant\footnote{2MASS J23154868-1050583} appearing in the DR3 survey is located at $\sim13"$, with a G band magnitude of $\sim20.8$, much fainter than that of \system, at $12$. We thus select the \texttt{OPTIMAL} aperture, which minimizes the RMS of the light curves.

To correct and detrend each CHEOPS light curves from systematic due to CHEOPS orbit or spacecraft effects, we use the \texttt{pycheops} package \citep{pycheops}. We mask points for which the photometry deviates by more than $1.5\times\mathrm{MAD}$ (mean absolute deviation). We also mask points where the flux is strongly affected by cosmic ray hits during passages through the South Atlantic Anomaly or when large deviations of centroids of the Point Spread Function (PSF) of the target are measured. To correct for non-astrophysical noise sources such as background or stray light from the roll angle, we detrend the light curve using the basis vectors provided by the DRP. For each visit, we follow the recommendation of the DRP for the decorelations flags to use. The detrending vectors are then passed as free parameters when fitting the transits with a normal distribution centered on 0 ($\mathcal{N}(0, 1)$). To detrend from stray light due nearby objects reflecting in the telescope (or "glint") we fit a spline against the roll angle of the telescope. This is done in \texttt{pycheops} using the \texttt{add\_glint} function, which is fitted using uniform distribution between 0 and 2 ($\mathcal{U}(0,2)$). The cutoff values, vectors, glint orders used, and the remaining number of points are detailed in Table~\ref{table:reduction_params}.

\renewcommand{\tabcolsep}{3pt} % Making it fit on the page
\begin{table*}
\caption{Decorrelation values used for each light curve.}
\centering

\begin{tabular}{l c c c c c l l r}
    \hline
    ID & Background flux & \multicolumn{2}{c}{Centroid $x$} & \multicolumn{2}{c}{Centroid $y$} & Decorrelation vectors & Glint order & Remaining \\
     & & min & max & min & max & & & points \\
    \hline\hline
    101 & 0.1572 & 262.8 & 264.0 & 839.15 & 840.2 & & $9^\dag$ & 181 \\
    102 & 0.17 & 264.75 & 265.7 & 840.6 & 841.35 & \emph{dfdt} & $17^\dag$ & 195 \\
    103 & 0.155 & 280.3 & 281.2 & 825.5 & 826.75 & & 15 & 186\\
    201 & 0.156 & 262.4 & 263.8 & 838.8 & 839.9 & \emph{dfdcos2phi} & 25 & 236 \\
    202 & 0.156 & 263.1 & 264.2 & 839.3 & 840.2 & \emph{dfdsmear} & $13^\dag$ & 154 \\
    203 & 0.156 & 279.8 & 280.8 & 825.3 & 826.5 & & $9^\dag$ & 130 \\
    301 & 0.145 & 264.8 & 265.6 & 839.9 & 840.8 & \emph{dfdcos2phi} & 15 & 148 \\
    302 & 0.147 & 264.0 & 264.9 & 839.5 & 840.6 & \emph{dfdsin2phi, dfdcos2phi} & 15 & 143 \\
    303 & 0.153 & 263.5 & 264.6 & 839.6 & 840.6 & \emph{d2fdt2, d2fdy2, dfdcos2phi} & 17 & 241 \\
    304 & 0.16 & 262.25 & 263.25 & 838.75 & 839.5 & \emph{dfdcos2phi} & 13 & 138 \\
    305 & 0.155 & 280.5 & 281.5 & 825.4 & 826.6 & & 13 & 156 \\
    501 & 0.179 & 279.0 & 281.0 & 824.6 & 826.4 & \emph{dfdt, dfdcosphi} & $23^\dag$ & 621, 628, 711 \\
    \hline
\end{tabular}
\tablefoot{\textdagger: Glint computed with respect to the moon position rather than roll angle. For visit 501, the three figures are for planets $d$, $f$ and $g$, respectively.}

\label{table:reduction_params}
\end{table*}

% Description

\cheops{} visits are typically setup to catch one transit of a single planet, and we scheduled multiple visits for planets $d$, $e$, and $f$ (table~\ref{table:cheops_files}). This way, we observed four transits of planet $d$ with \cheops{}, thus extending the transit monitoring up to $250$ orbits after the first epoch, and three transits of planet $e$. The ephemeris for planet $f$ predicted the transit $\sim2.5$ hours too late. Combined with the $3.2$ hours transit duration, three visits completely miss the first half of the transit. Additionally, the egress falls in the observation gaps of the \cheops{} orbits in four visits. In total, out of the six available visits, only two egresses are perfectly caught. Because of the uncertainty on the previous ephemerides, some transits were predicted to occur in the \cheops{} visits, but were not detected. We identified those in the first column of Table~\ref{table:cheops_files}. Notably, planets $b$ and $c$ were not detected at all, while only a single transit of planet $d$ was not detected.

Given the number of planets in this system and their short orbital periods, the ephemerides from HU21 and A22 predict light curves with multiple transits. Five visits should contain two or more transits of the different planets in the system: 102 ($b$, $d$), 201 ($b$, $c$, $e$), 303 ($d$, $f$), 304 ($b$, $f$), 501 ($b$, $d$, $f$, $g$) (see Figs.~\ref{fig:4_transit},~\ref{fig:missing_b}, and~~\ref{fig:planet_c_201}).

\subsection{Modeling of the \cheops{} light curves}
\label{modelling}

We perform the \cheops{} light curves analysis with the \texttt{pycheops} package \citep{pycheops}, routinely used for \cheops{} observations analyses \citep[e.g.][]{TOI-178-Delrez, Hoyer_2023}. This is done in two steps. A first estimate of the model parameters are determined using \texttt{LMFIT}\footnote{\url{https://lmfit.github.io/lmfit-py/index.html}} \citep{LMFIT}, which fits the transit using a least-square scheme. Then the parameters are refined using the \texttt{emcee} package \citep{EMCEE}, which explores the parameter space using an MCMC scheme. This second step allows for a better estimation of the fit, as well as the errors associated to each parameter. We use a normal distribution on planetary and stellar parameters priors based on their mean and $1\sigma$ interval values from A22, given in Tables~\ref{table:stellar_parameters}~and \ref{table:planetary_parameters}.

Unlike most \cheops{} visits that are of short duration, aiming to capture unique transits, visit 501 exhibits four transit-like features (Fig.~\ref{fig:4_transit}). We first identified the planets most likely to be associated with each transit observed in this visit. Based on the ephemerides from A22, we fitted each of the observed planets in their respective visits to get a good estimate of their transits timing. We have used these updated ephemerides when possible (and retained those of A22 in cases where they were not available) to determine which planets were the most likely observed during this visit. We thus identified the first three transits as being due to planets $d$, $f,$ and $g$, respectively. Planet $b$ appeared as a the most likely candidate to explain the fourth transit feature in this visit. However, as is detailed in Sect.~\ref{ssec:planet_b}, we finally discarded this possibility.

The \texttt{pycheops} package is designed so that only one transit per light curve can be analyzed. To overcome this limitation for visit 501, to fit each transit event, we mask the other unwanted events using the predicted positions of the planets. Therefore, we were able to proceed with the fit as if a single transit was present in each case. This allowed us to follow the same procedure as described above for each transit. \texttt{pycheops} also allows also for multi-visit modeling \citep[e.g.][]{Hoyer_2023}. Transits of planet $d$, $f,$ and $g$ of visit 501, were included in their respective multi visit analysis. The resulting fits on visit 501, including the putative planet $b$, are shown with their ephemerides in Fig.~\ref{fig:4_transit}.

For each planet, we detailed below the results obtained  from the analysis of our \system{} \cheops{} observations (Sect.~\ref{ssec:cheops_data}). The results of the \texttt{pycheops} fits, solely based on \cheops{} data, are presented in Table~\ref{table:planetary_parameters}.

%--------------------------------------------------------------------
% Section
%--------------------------------------------------------------------

\section{\cheops{} Transits analysis}\label{sec:results}

%Note that while the values are presented to show the results of the \texttt{pycheops} analysis, only the individual transit mid-times (listed in Table~\ref{tab:ttvs}~\&~\ref{tab:ttvs_continuacion}) are used when computing the new ephemerides.

\subsection{Planet $b$}\label{ssec:planet_b}

\begin{figure}
    \centering
    \includegraphics[width=\linewidth]{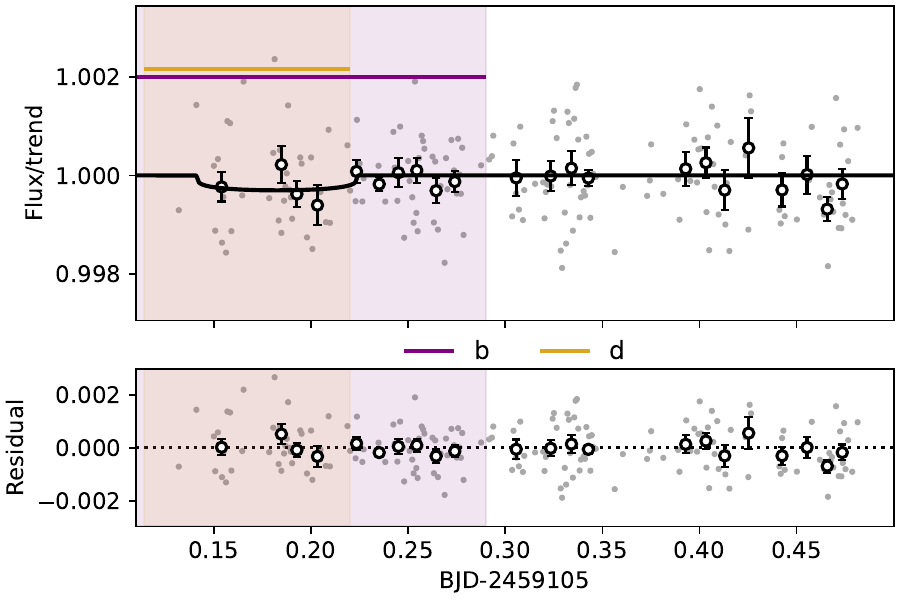}
    \includegraphics[width=\linewidth]{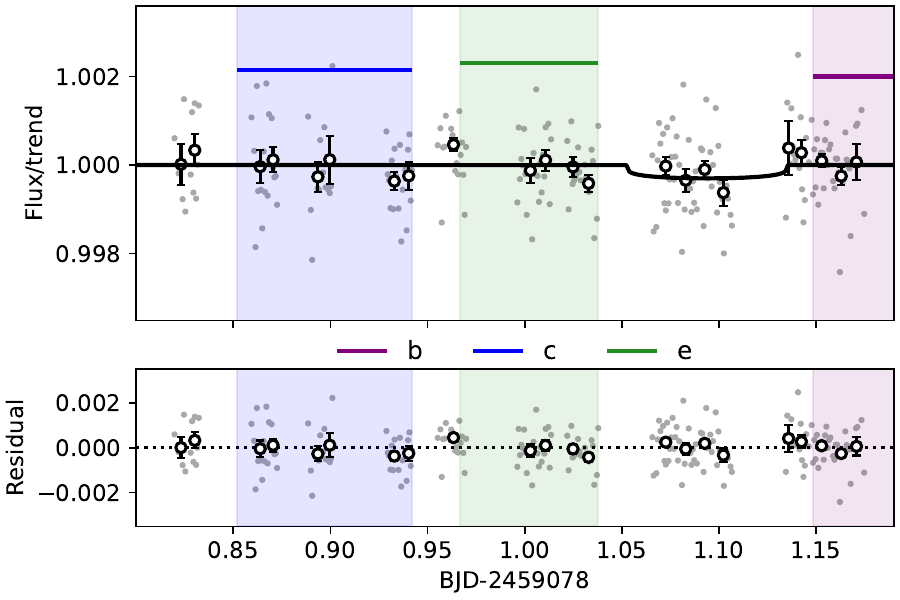}
    \includegraphics[width=\linewidth]{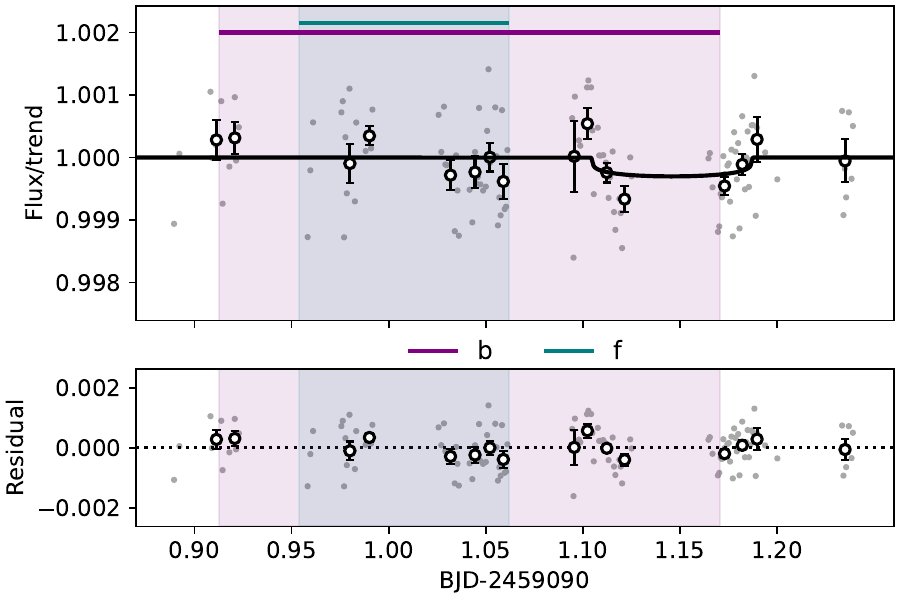}
    \caption{Putative transits of planet $b$. From top to bottom: Light curves of \cheops{} visits 102, 201 and 304, in which planet $b$ transits were expected to occur. Colored regions show the expected transits mid-time with their $\pm 1 \sigma$ errors. Transits due to other planets than $b$ have already been removed. The solid black curve represent the best fits for planet b, while fixing its depth and duration to $D = 252$\,ppm and $T_{14} = 2.00$\,h.}
    \label{fig:missing_b}
\end{figure}

With an expected depth of around 250~ppm (see Table~\ref{table:planetary_parameters} for details), the transit of planet $b$ is at the edge of detectability with \cheops. Since the last observed transit of planet $b$ dates back to \kt{} and that its current ephemeris predict that it should be visible in four of our visits, we nevertheless attempted to find it in our dataset. No visit was aimed directly at planet $b$, but it should instead appear on visits targeting other planets (see Table~\ref{table:cheops_files}). We show the expected transits (based on A22's ephemeris) in Figs.~\ref{fig:4_transit}~and~\ref{fig:missing_b}. In the cases of visits 201 and 501, the predicted transit time and its $1\sigma$ error is fully contained within the observations, unlike visits 102 and 304.

Since visits 102, 201 and 304 are quite short due to focusing on a single planet, we attempted to fit planet $b$'s transit signature in the residual of each visit (Fig.~\ref{fig:missing_b}). We found no clear evidence of a photometric dip corresponding undoubtedly to planet $b$,  therefore, we do not use them in the following TTV analysis. Visit 501 was aimed at capturing a transit of planet $g$ and is a much longer observation than a typical \cheops{} visit. We therefore expected good chances of seeing a transit of planet $b$ in this dataset. The best case scenario obtained is the one shown in Fig~\ref{fig:4_transit}, where a transit-like feature is observed almost concomitantly to the predicted transit of the planet. However, the fitted transit values are not compatible with those of the literature. The modeled transit has indeed a duration $T_{14,\mathrm{fit}} = 1.12 \pm 0.08$\,h compared to the expected $T_{14,\mathrm{prior}} = 2.00^{+0.09}_{-0.11}$\,h, as well as a depth of $D_\mathrm{fit} = 557 \pm 160$\,ppm to $D_\mathrm{prior} = 252^{+23}_{-21}$\,ppm. Because of this mismatch on the duration and on the depth, we do not consider it to be a transit of planet $b$. This short event could be the residuals of some systematics during the detrending.

\subsection{Planet $c$}\label{ssec:planet_c}

\begin{figure}
    \centering
    \includegraphics[width=\linewidth]{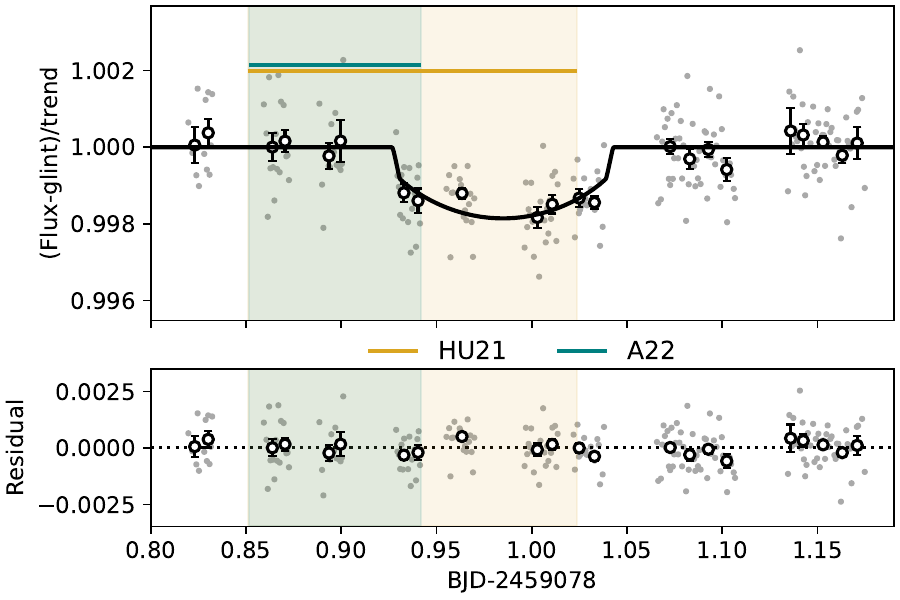}
    \caption{Detrended light curve of \cheops{} visit 201. The solid black curve represents the fitted transit of planet $e$. The predicted mid-transit time of planet $c$ is shown within $\pm 1 \sigma$ error as colored regions from two different sources: A22 in green and HU21 in yellow.}
    \label{fig:planet_c_201}
\end{figure}

We expected to catch the planet $c$ transit in visit 201, as shown in Fig.~\ref{fig:planet_c_201}, together with those of planets $e$ and $b$. With an estimated depth of $\sim500$~ppm, planet $c$ should be detectable with \cheops{}. Unfortunately, this visit was strongly affected by gaps and the only transit-like feature appears outside the expected ranges for the three planets (Fig.~\ref{fig:missing_b}). Given the uncertainties in planets $c$ and $e$ ephemeris, however, it remains close to these ranges. We therefore explore the possibility that the transits of the two planets overlap, the event been masked by the numerous interruptions in the light curve that would hide clear ingress and egress of transits.

\begin{figure*}
    \centering
    \includegraphics[width=0.47\linewidth]{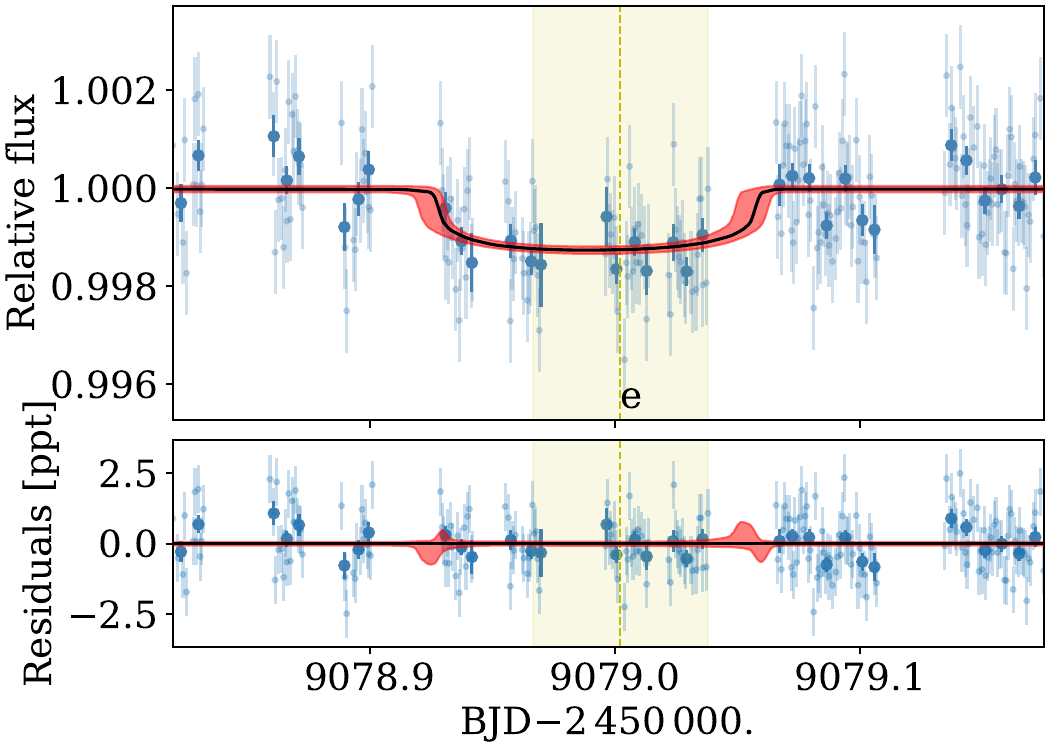} 
    \includegraphics[width=0.47\linewidth]{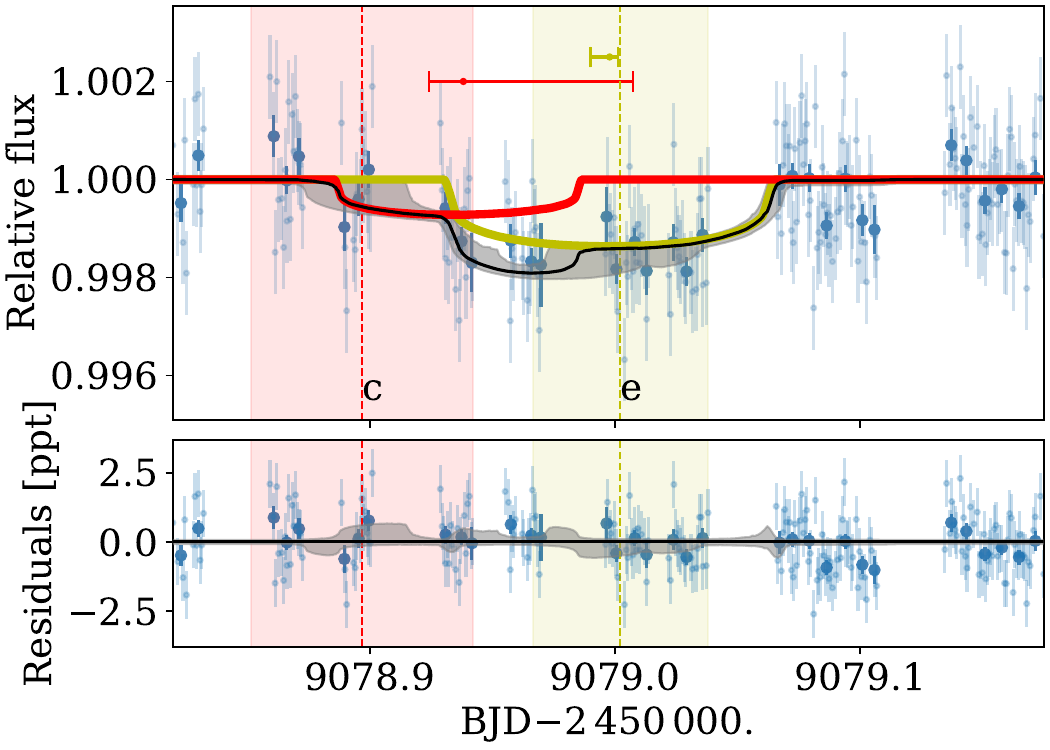}
    \caption{Modeling of \cheops{} visit 201. The detrended photometric points and its ten-minute bins are represented by the light and dark blue symbols. The predicted transit times from A22 ($\pm1\sigma$) of each planet are represented by the vertical regions. \lang{The left} panel shows the fit of planet $e$ transit only. The model and its $\pm1\sigma$ uncertainties shown with the black curve and red region, respectively. The right panel shows the fit including transit of both planets $c$ and $e$. The fitted transit of planet $c$ and $e$ are represented by the red and yellow curves, respectively, while the full model ($\pm1\sigma$) is represented by the black curve (and gray area). The two horizontal lines at the top of the panel represents the fitted transit mid-time for each planet and its $\pm1\sigma$ errors. }
    \label{fig:visit_201}
\end{figure*}

To that end, we performed a one-planet (planet $e$), and two-planet (planets $e$ and $c$)  transit-fitting modeling successively, using the \texttt{juliet} code \citep{juliet}. The advantage of \texttt{juliet}, is that it allows the simultaneous fit of two planetary transits and also a straightforward comparison of the results based on Bayesian criteria. For these fits (apart from the transit mid-time), we fixed all the parameters to the central values of planets $c$ and $e$ priors reported in Table~\ref{table:planetary_parameters}. For the transit mid-time of planet $e$ we used a normal distribution defined by BJD$-$2\,450\,000 $=9078.98\pm0.3$. For the mid-time of planet $c$ we defined an uniform distribution between BJD$-$2\,450\,000 $= 9078.905$ -- 9079.065, which was the wider distribution possible that secured the convergence of the fit. 
In Fig.~\ref{fig:visit_201}, we present both solutions and the corresponding mid-times are given in Table~\ref{table:planetary_parameters}. As expected, the mid-times of planet $e$ obtained from both models are consistent. The fitted mid-time of planet $c$ ranges along the first half of the transit of planet $e$, namely, between the ephemeris' predicted time (as marked by the red vertical region in right panel of Fig.~\ref{fig:visit_201}) up to the middle of planet $e$ transit. To determine which is the best model, \texttt{juliet} provides the Bayesian evidences ($\ln(Z)$) of each fitted scenarii (see Table~\ref{tab:visit_201_juliet}). In this comparison, the planet-$e$ only model yields a log-evidence of $\ln(Z)=1208.59 \pm 0.46$, while that of the planets-$e$ and $c$ model is of $\ln(Z)=1206.37 \pm 0.44$, leading to a $\Delta\ln(Z)=2.2 \pm 0.7$. The limit between weak and moderate evidence of preference being $\ln(Z)=2$, we can therefore not statistically prefer one model over the other \citep{juliet}. This situation is discussed further when we compute the new ephemerides in in Sect.~\ref{sec:ephemerides}.

\renewcommand{\arraystretch}{1.2} % Better spacing for readability
\begin{table}[]
\caption{Transit mid-times and their Bayesian evidence of planet $e$ and $c$ in visit 201 light curve.
%These times were obtained with the fit of one- and two-planet models with \texttt{juliet}.
}
    \label{tab:visit_201_juliet}
    \centering
    \begin{tabular}{cccc}
    \hline
    \hline
    Model & Planet &T$_{\mathrm{c}}$ [BJD - 2\,450\,000] & $\ln(Z)$\\
    \hline
    1-planet &     $e$   & $9078.9927^{0.0039}_{0.0081}$ & $1208.59 \pm 0.46$\\
    2-planet  &  $e$ &  $9078.9978^{0.0035}_{0.0078}$ & $1206.37 \pm 0.44$\\
    & $c$ & $9078.938^{0.069}_{0.014}$ & \\
    \hline
    \end{tabular}
\end{table}
\renewcommand{\arraystretch}{1.0} % Better spacing for readability

\subsection{Planet $d$}\label{ssec:planet_d}

Transits of planet $d$ are present in visits 101, 102, 103, and 501. The ephemeris for this planet from A22 predicts the transit roughly 3.5 hours too early in each case, but the \cheops{} observations still manage to capture four complete transits (see Fig.~\ref{fig:d-transit}). Three of the visits catch the ingress of the transit, while one shows a partial egress. While we were unable to catch both features on a single visit, we are able to derive accurate transit mid-times to use in our subsequent TTV analysis. Notably, planet $d$ is also predicted in visit 303, but not detected. This is likely due to the short duration of the visit, which prevented us from catching the transit.

\subsection{Planets $e$}
\label{ssec:planets_e}

This planet had the most accurate predicted transit times among all the planets in the system, with all its transits occurring within the estimated error range. Although visit 201 may have been affected by the potential transits of planets $b$ and $c$, the other two visits are not anticipated to show any additional transits. However, due primarily to insufficient coverage during the ingress and egress phases of the transit, we faced limitations in the precision of the obtained transit mid-times (see Fig.~\ref{fig:e-transit}).

\subsection{Planet $f$}
\label{ssec:planet_f}

Planet $f$ appears most frequently in our dataset, being observed during visits 301, 302, 303, 304, 305, and 501. The first five visits were dedicated to observing this planet, while visit 501 targeted planet $g$.
%Although planet $b$ was expected to be detected during visit 304, no signal from that planet was observed (see Sect.~\ref{ssec:planet_b}), planet $d$ is also not detected in visit 303.
Despite the number of visits dedicated to observing planet $f$, the ephemeris used to prepare these observations deviated significantly as illustrated in Fig.~\ref{fig:f-transit}. This results in predicted transit times later than observed. Consequently, we captured only one egress of the transit during visit 501. This discrepancy led to mid-transit times with deviations from a constant orbital period of up to $\sim25$ minutes, with an average error of $\sim11$ minutes, which allows for a TTV analysis of the system to be carried out.

\subsection{Planet $g$}
\label{ssec:planet_g}

When planning observations with \cheops{}, only two transits of the planet with a 42-day orbit had been observed with \kt{} \citep{christiansen2018, lopez2019}, and an additional transit was captured with \spitzer{} (HU21). However, the projected uncertainties of the transit times were significantly larger than 1.6 hours based on these observations. To address this, the visit for planet $g$ was extended beyond the usual duration of \cheops{} visits, allowing us to also capture transits of planets $d$ and $f$ (see Sects.~\ref{ssec:planet_d}~and~\ref{ssec:planet_f}). After the transit modeling, we were able to determine the mid-transit time of planet $g$ with a precision of approximately $\sim$12.25 minutes at epoch 32. This extends the observation coverage of the planet beyond its last observation at epoch 10 by \spitzer{}. Such comprehensive coverage is essential for estimating TTVs in the system, therefore, we have included this transit in our analysis.

%--------------------------------------------------------------------
% Ephemerides
%--------------------------------------------------------------------

\section{Updated ephemerides}\label{sec:ephemerides}

\renewcommand{\arraystretch}{1.2}
\begin{table}[]
    \caption{New ephemerides for \system.}
    \label{tab:new_ephemeris}
    \centering
    \begin{tabular}{c c c}
    \hline
    Planet & Average period [days] & $T_0$ [BJD - 2\,450\,000] \\
    \hline\hline
    $c$\supdagger & $3.559823 \pm 0.000050$ & $7740.3217 \pm 0.0003$ \\
    $d$ & $5.405362 \pm 0.000038$ & $8791.7962 \pm 0.0054$ \\
    $e$ & $8.261334 \pm 0.000011$ & $7740.6459 \pm 0.0007$ \\
    $f$ & $12.756618 \pm 0.000061$ & $7738.7053 \pm 0.0051$ \\
    $g$ & $41.965443 \pm 0.000061$ & $7773.8552 \pm 0.0016$ \\
    \hline
    \end{tabular}
    \tablefoot{\textdagger: Estimated using only \kt{} data.}
\end{table}
\renewcommand{\arraystretch}{1.0}

The new ephemerides of the planetary transits of this system make use of the \kt, \spitzer, the \hale{} telescope, and the newly acquired \cheops{} data, including the transits from planets $d$, $e$, $f,$ and $g$. The \kt{} data and their respective mid-transit times were first used in \citet{christiansen2018} and \citet{lopez2019}, but unfortunately, the individual transit mid-times were not reported by the authors. Thus, we reproduced the procedure used in both studies: the individual \kt{} transit mid-times were measured by fixing the model transit parameters to the best-fit values given in previous studies \citep{christiansen2018, lopez2019, hardegree-ullman2021, boyle2022} and allowing only the mid-transit time to vary. The uncertainties were calculated by computing the residuals from the best-fit model and performing a bootstrap analysis using the closest 100 timestamps, re-fitting the mid-transit time at each timestamp permutation. All the mid-times are listed in Table~\ref{tab:ttvs}~and~\ref{tab:ttvs_continuacion} for each planet. For the rest of the data (\spitzer\ and \hale), the individual transit mid-times were provided directly by the respective authors, therefore, there was no need for a reanalysis.

With all the available transit times of each planet, we built the respective \lang{observed minus calculated} (O$-$C) diagrams of the transits (Figs.~\ref{fig:OC_c} to \ref{fig:OC_g}), where we compared the observed transit times with our updated ephemeris as described below. For each planet, the updated values of the average period and reference epoch ($T_0$) were obtained by a weighted linear fit of the transit times. The updated values of the average periods and $T_0$ are given in Table~\ref{tab:new_ephemeris}. The time residuals are also listed in Table~\ref{tab:ttvs} and \ref{tab:ttvs_continuacion}. The reported errors of the time deviations consider the uncertainty of the updated ephemeris and the respective error of the observed mid-time.

To improve the ephemeris of planet $c$, we utilized the transit times from \kt{} and three reference epochs (T$_0$) provided in \citet{lopez2019}, HU21 and A22 designated as epoch 0. All transit times are listed in Table~\ref{tab:ttvs}. If the mid-time of the potential \cheops{} transit in visit 201 is excluded, fitting the transit times with a linear function yields an updated ephemeris that diverges notably from those of HU21 and A22. This discrepancy is illustrated in Fig.~\ref{fig:OC_c}, which presents the O$-$C diagram of planet $c$'s transit times using our updated ephemeris. Our refined ephemeris aligns with the scenario of a non-detection during \cheops{} visit 201, as the planet was likely to have transited shortly before or at the very beginning of the light curve. Conversely, as previously discussed and as shown in Fig.~\ref{fig:OC_c}, assuming the detection of the transit in visit 201 results in a fitted time consistent with the predictions of HU21 ephemeris. Given the inability to definitively rule out any of these scenarios with our current data, any future transit monitoring of this planet ought to consider the large uncertainties in transit time predictions.

The ephemerides of planets $d$ and $e$ were refined using all available previous data in addition to observations from \cheops{}, as detailed in Table~\ref{tab:ttvs} and ~\ref{tab:ttvs_continuacion} respectively. For comparison, we also show in their updated O$-$C diagrams (Figs.~\ref{fig:OC_d_v2} and \ref{fig:OC_e}) the time projections from previous transit ephemerides. For planet $d$, our \cheops{} times do not agree with the linear ephemeris defined by considering only the \kt{} and B22 transits. When incorporating the \cheops{} data, the B22 transit times present a significant deviation of $-158\pm14$\,min and $-48\pm12$\,min for planets $d$ and $e$, respectively, with respect to our updated linear ephemeris. This result is discussed in the next Sect.~\ref{ssec:ttv_modeling}.

The six new \cheops{} visits for planet $f$ effectively double its number of observations. Despite the incomplete coverage during these visits (as discussed in Sect.~\ref{ssec:planet_f}), we were able to improve the ephemeris of the planet. As illustrated in Fig.~\ref{fig:f-transit}, the ephemeris from A22 exhibited drift, resulting in predictions that were consistently too early by $121\pm76$ to $164\pm77$ minutes.

With only three previous observations, the ephemeris of planet $g$  were not very well constrained with projected uncertainties of hours, as illustrated in Fig.~\ref{fig:OC_g}. The new \cheops{} observation allows us to correct the ephemeris: it accounts for a deviation of around 45 minutes with respect to HU21 ephemeris for the \cheops's epoch and also constrains the uncertainties of the projected transit times down to only few minutes (see Fig.~\ref{fig:OC_g} and Table~\ref{tab:new_ephemeris}).

%--------------------------------------------------------------------
% Modeling TTVs
%--------------------------------------------------------------------

\section{TTV analysis }\label{ssec:ttv_modeling}

Despite the fact that the number of transits observed are not enough to perform a full TTV analysis of the system and thereby enable a probe of the masses and/or eccentricities distribution of planets using photometry alone \citep[e.g.,][]{TOI-178-Leleu, HD108236_Hoyer2022}, we conducted a first-order TTV analysis. To that end, we used the \texttt{TTVfaster}\footnote{\url{https://github.com/ericagol/TTVFaster/}} code \citep{AgolDeck2016} for the TTV simulations combined with the nested sampling method \citep{Buchner2021a}. This code implements analytical formulas instead of numerical integrations to model the TTVs of multi-planetary systems, allowing for more efficient transit timing modeling of the planets in packed systems. This approach helps us constrain the expected TTVs and validate our updated orbital periods as reported in Table~\ref{tab:new_ephemeris}. Such modeling is particularly crucial, for example, in verifying the significant TTVs observed in planet $d$.

Our approach is well suited when exploring non-Gaussian parameters spaces. This is expected in our case, as we have a large number of fitted parameter compared to the number of transit times we have. Indeed \texttt{TTVfaster} requires seven input parameters per planet. For this we use the implementation of the \texttt{UltraNest}\footnote{\url{https://johannesbuchner.github.io/UltraNest/}} package \citep{Buchner2021b},  in particular, its \texttt{stepsampler.SliceSampler} method, which is more efficient for high-dimensional spaces. We analyzed the TTVs of the planets based on the timing information detailed in Sects.~\ref{sec:results}~and~\ref{sec:ephemerides} and reported in Tables~\ref{tab:ttvs}~and~\ref{tab:ttvs_continuacion}, combined with the masses values from A22. In addition, we used  our updated planetary parameters as priors. With respect to the parameters we lacked updates for, such as those of planets $b$ and $c$, we utilized the values reported in A22. Specifically, for planet $b$, we relied on the three time stamps provided by A22, as we have no significant detection of any transit of this planet; thus, we were not able to update its ephemeris. Therefore, little information from this planet can be used to constrain the planet-system parameters. The defined priors distribution for each parameter, as well as the posteriors, are detailed in Table~\ref{tab:ttvfaster}. We report in this table the distribution of the relevant fitted planetary parameters: mean orbital period (P), mass (M), and eccentricity (e). In addition, the RMS and the larger amplitude of the best-fit TTV model are also reported.

\begin{figure*}
    \centering
    \includegraphics[width=0.45\linewidth]{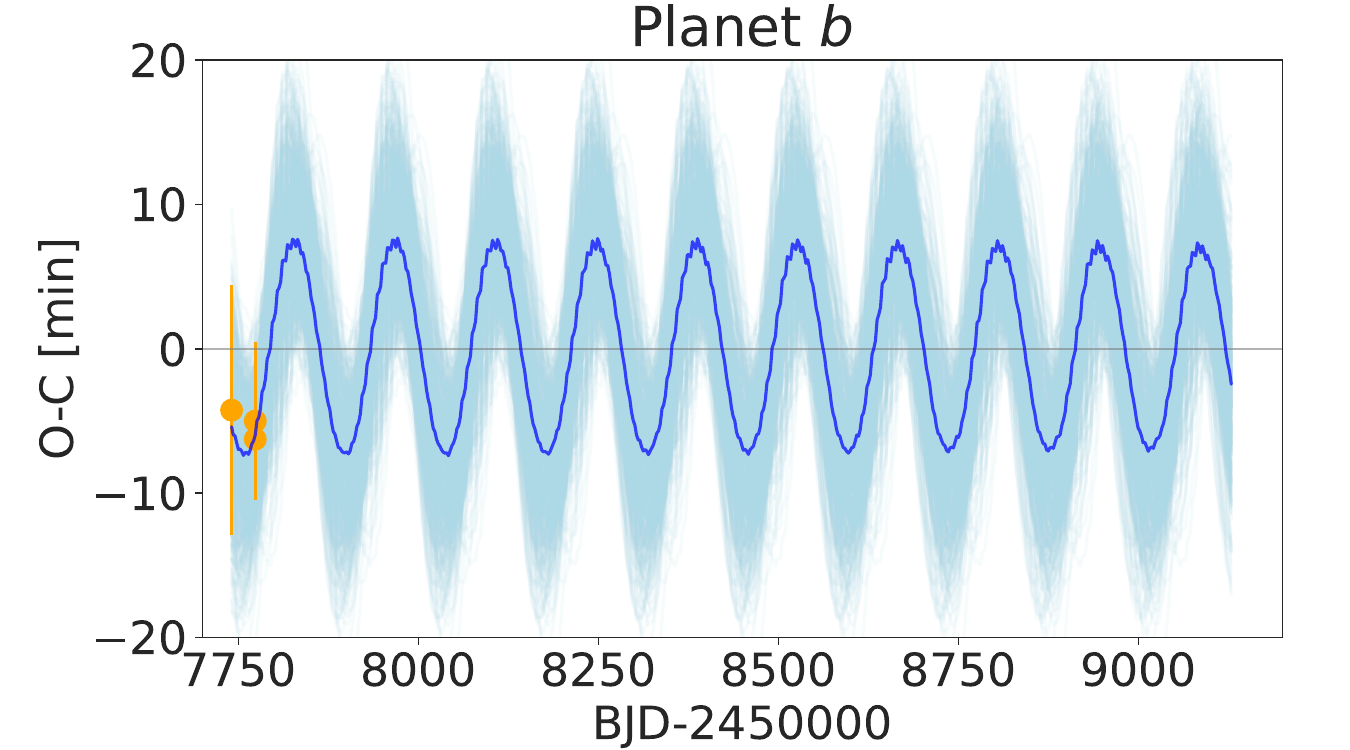}
    \includegraphics[width=0.45\linewidth]{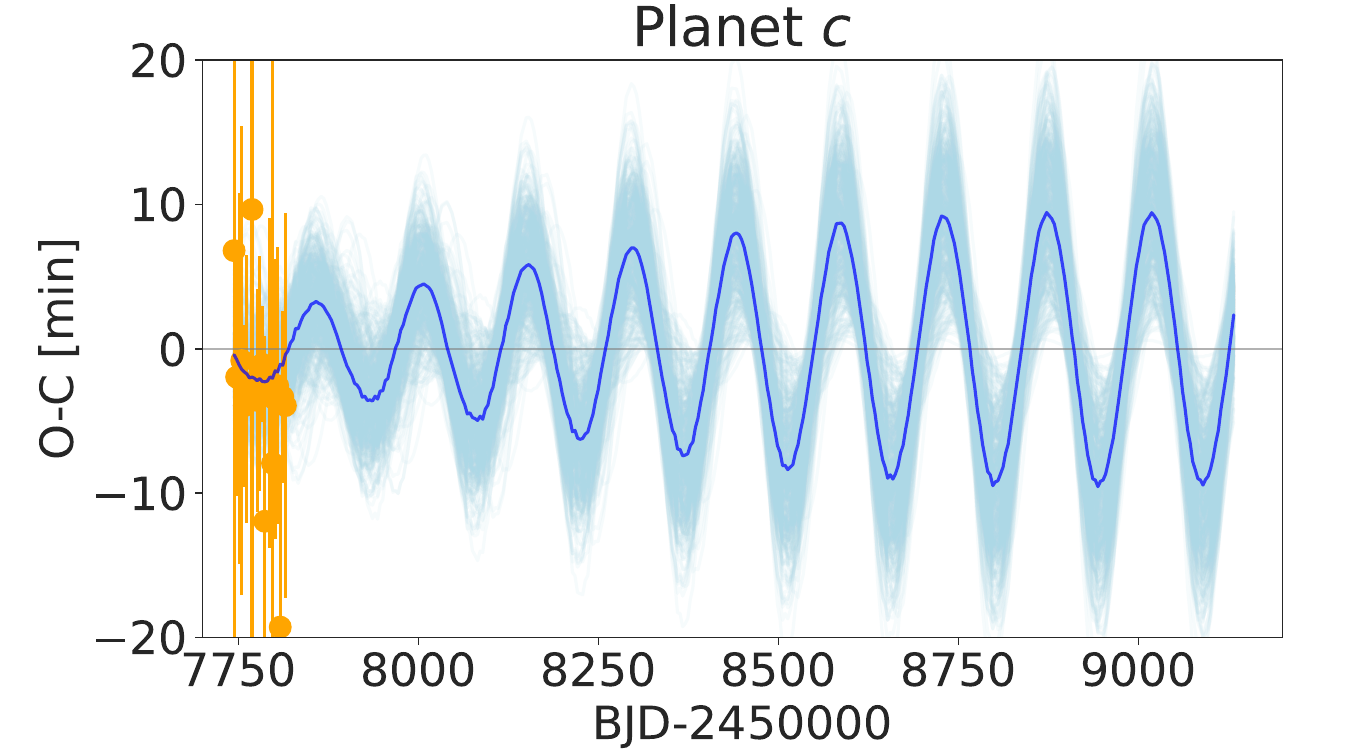}\\
    \includegraphics[width=0.45\linewidth]{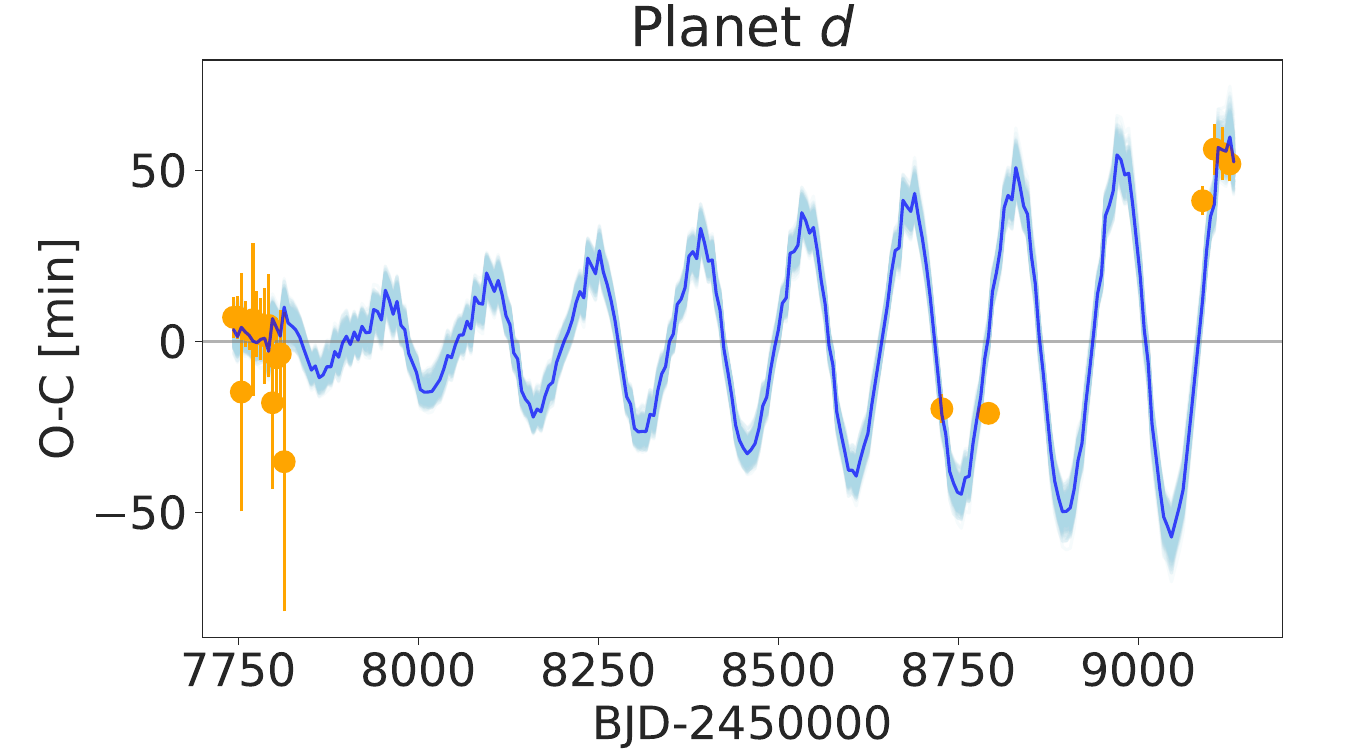}
    \includegraphics[width=0.45\linewidth]{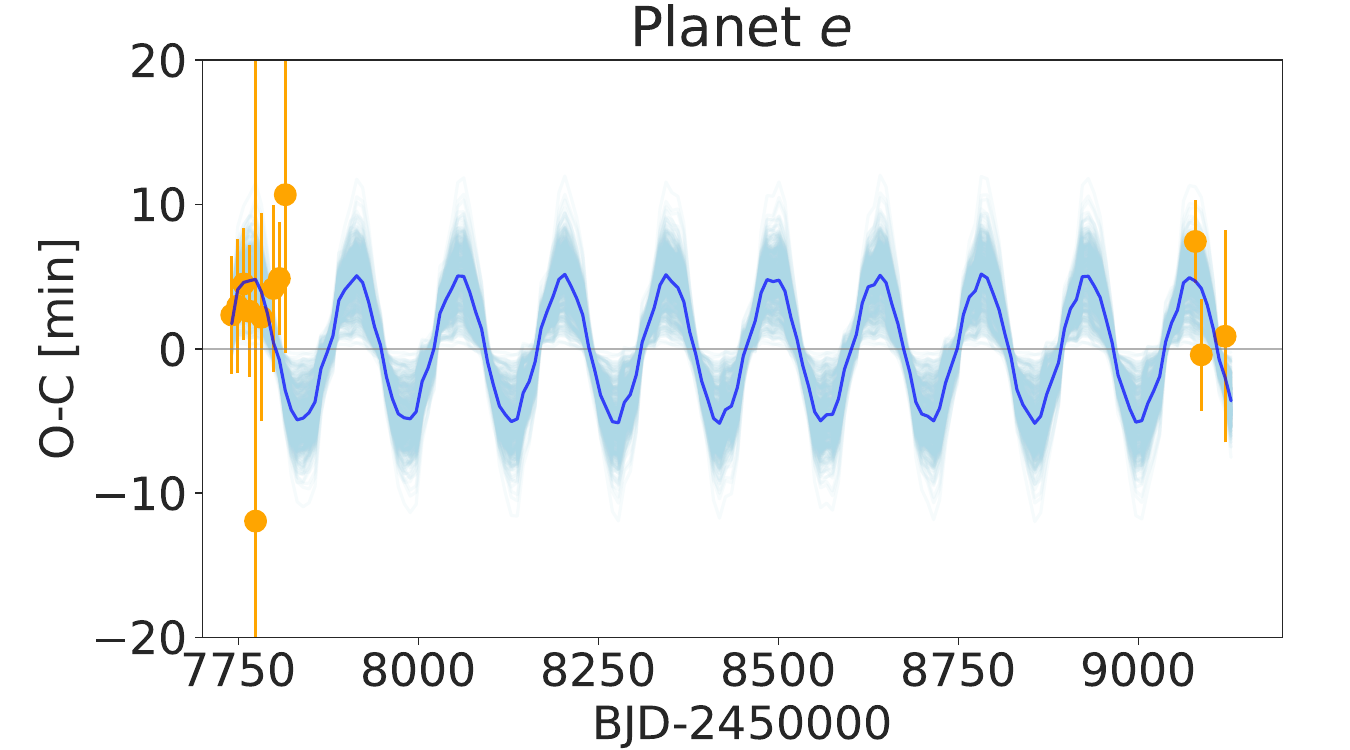}\\
    \includegraphics[width=0.45\linewidth]{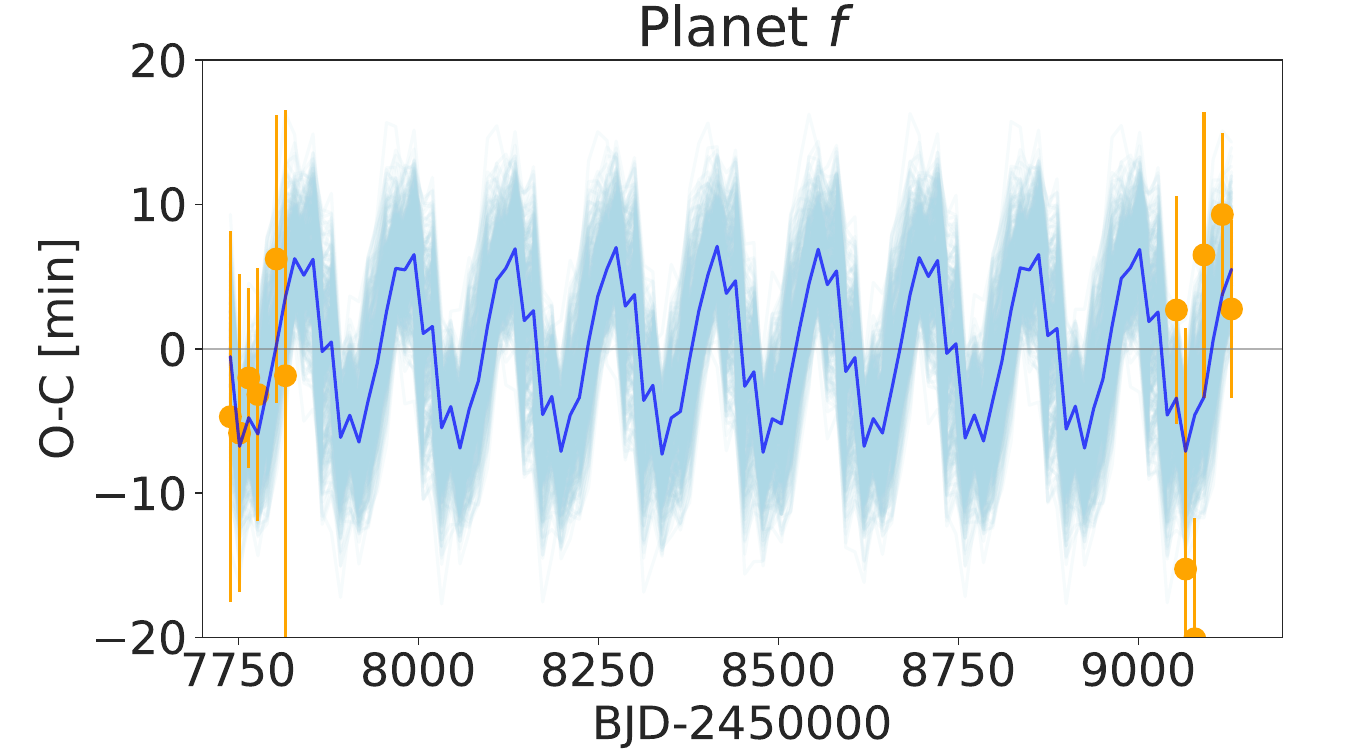}
    \includegraphics[width=0.45\linewidth]{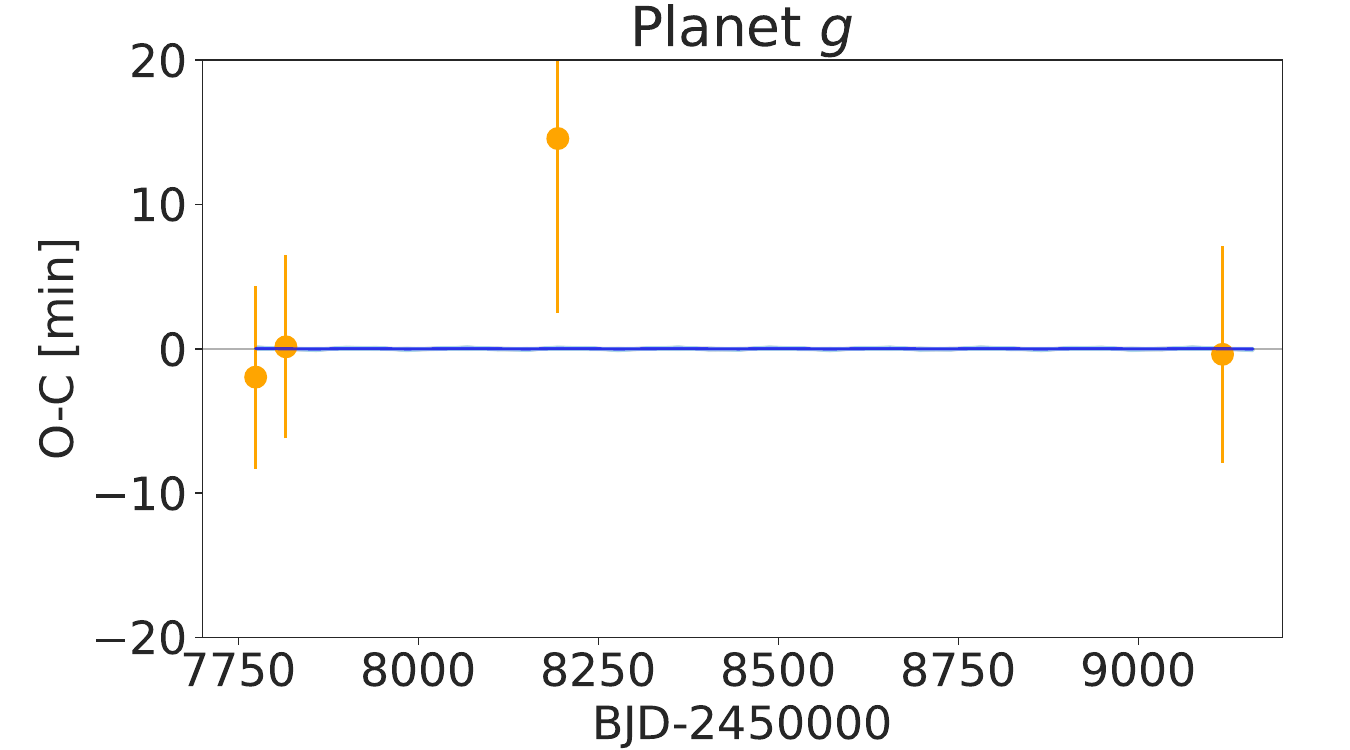}
    
    \caption{Results of TTV simulations of the \system\ planetary system. The orange points represent the observed transit time residuals for each planet, reported in Tables~\ref{tab:ttvs}~and~\ref{tab:ttvs_continuacion}. The best-fit TTV models obtained with \texttt{TTVfaster} and its $\pm1\sigma$ errors are represented respectively by the blue curve and region, and are detailed in Table~\ref{tab:ttvfaster}.}
    \label{fig:ttv_sims}
\end{figure*}

The time residuals of the planetary system simulation are presented in the O--C diagrams of Fig.~\ref{fig:ttv_sims}, along with the best-fit TTVs model. The posterior values of the fitted parameters are reported in Table~\ref{tab:ttvfaster}. As also reported in the table, the models predict mean TTV amplitudes that are lower than 10~minutes for planets $b$, $c$, $e$, $f,$ and $g$, as depicted in Fig.~\ref{fig:ttv_sims}.  For planet $d,$ however, the temporal deviations are as large as 60 minutes, with a modeled TTVs RMS of 25 minutes. The fitted planetary parameters of the modeled TTVs are fully consistent with the defined priors, particularly the mean orbital period, eccentricities, and masses. Planet $d$ is the only notorious exception, where its fitted mass deviates to a small $M_{d}\sim0.8$\mearth, and a relatively high eccentricity of $e_{d}=0.16\pm0.04$. These values were necessary to fit the large TTVs observed with \cheops.

It is clear that the transit times can not be described purely with a strict constant period. Notably for planet $d$, as reported in Table~\ref{tab:ttvs}, B22's transits deviate $-158\pm14$\,min and $-48\pm12$\,min from the updated ephemeris, while three out of the four \cheops{} transits deviate by more than 10 minutes (but with uncertainties that are on the same order). Assuming these deviations are a consequence of gravitational interactions between the planetary bodies, we can predict that the actual mean orbital period of planet $d$ should therefore be between the B22 value and our updated estimate. This detection of large TTV indicates the significant dynamical interaction of planet $d$ with the other planets in the system, as demonstrated by our simulations. New observations would be required to constrain the presence of these TTVs, which are expected for planets located in MMR \citep[e.g.,][]{agol2005, Xie2013}, as is the case of planet $d$.% (in a resonance chain of 3:2 with planet $c$ and $e$).
These observations would also be required to properly sample the super-period of planet $d$, and further constrain its eccentricity.

For planet $f$, the time residuals shown in Fig.~\ref{fig:OC_f} seem to suggest a hint of TTVs. In fact, the \cheops{} transits show a more scattered distribution around the new linear ephemeris than the \kt{} points, though the time uncertainties prevent us to claim this is a significant TTV detection. Moreover, the TTV simulations only predict small periodic oscillations with amplitudes of 7.4 $\pm$ 3.2 minutes around a constant period.  Thus, more precise measurements are required to confirm the presence of such deviations.
No deviations were predicted from the simulations for planet $g$ as it is
located further out  in the system; therefore, it is subject to only \lang{weak} gravitational influence from the innermost planets.

\begin{table*}[]
\caption{Priors and posterior distributions of the TTV modeling.}
\label{tab:ttvfaster}
    \centering
    \small
    \begin{tabular}{l c c c}
    \hline
    \hline
      Parameter   &  Prior  & Priors values & Posterior \\
                  &  distribution & ($\mu$,$\sigma$) & \\
     \hline
     P$_{b}$ [d] & $\mathcal{TN}$ & (2.353345, 0.000027)  &  $2.353349 \pm 0.000040$ \\
     P$_{c}$ [d]& $\mathcal{TN}$ & (3.5600, 0.00012)   &    $3.560027 \pm 0.000097$ \\
     P$_{d}$ [d]& $\mathcal{TN}$ & (5.40531, 0.000041)  &   $5.405244 \pm 0.000012$ \\
     P$_{e}$ [d]& $\mathcal{TN}$ & (8.2613550, 0.0000075)  &   $8.261352 \pm 0.000008$  \\
     P$_{f}$ [d]& $\mathcal{TN}$ & (12.756683, 0.000034)  &  $12.756661 \pm 0.000029$  \\
     P$_{g}$ [d]& $\mathcal{TN}$ & (41.96501, 0.00017)  &   $41.96499 \pm 0.00016$ \\
     \hline
     
     M$_{b}$ [M$_{\oplus}$] & $\mathcal{TN}$ & (2.80, 0.96)  & $2.54\pm0.83$   \\
     M$_{c}$ [M$_{\oplus}$] & $\mathcal{TN}$ & (5.95, 1.18)  & $6.23\pm0.81$  \\
     M$_{d}$ [M$_{\oplus}$] & $\mathcal{TN}$ & (7.20, 1.41)  & $0.78\pm0.36$   \\
     M$_{e}$ [M$_{\oplus}$] & $\mathcal{TN}$ & (11.28, 2.79) & $14.11\pm 1.64$   \\
     M$_{f}$ [M$_{\oplus}$] & $\mathcal{TN}$ & (2.43, 2.40)  & $2.47 \pm 1.63$   \\
     M$_{g}$ [M$_{\oplus}$] & $\mathcal{TN}$ & (2.45, 2.33)  & $2.56\pm1.59$   \\
     
     \hline
    $e_{b}$ & h-$\mathcal{N}$ &(0.0, 0.1)  & 0.020 $\pm$ 0.015   \\
    $e_{c}$ & h-$\mathcal{N}$ &(0.0, 0.1)  & 0.015 $\pm$ 0.012   \\
    $e_{d}$ & h-$\mathcal{N}$ &(0.0, 0.1)  & 0.16 $\pm$ 0.04  \\
    $e_{e}$ & h-$\mathcal{N}$ &(0.0, 0.1)  & 0.014 $\pm$ 0.011   \\
    $e_{f}$ & h-$\mathcal{N}$ &(0.0, 0.1)  & 0.013 $\pm$ 0.010   \\
    $e_{g}$ & h-$\mathcal{N}$ &(0.0, 0.1)  & 0.09 $\pm$ 0.06   \\
     
     \hline
     \hline
     Planet &  TTV RMS & TTV amplitude  \\
     & [min] & [min] \\
      \hline
      $b$ & 5.2 & 7.7 $\pm$ 4.4 \\
      $c$ & 5.1 & 9.4 $\pm$ 2.8 \\
      $d$ & 25.1 & 60.0 $\pm$ 2.6 \\
      $e$ & 3.4 & 5.1 $\pm$ 1.3 \\
      $f$ & 4.5 & 7.4 $\pm$ 3.2 \\
      $g$ & 0.01 & 0.02 $\pm$ 0.02 \\
      \hline
    \end{tabular}
    \tablefoot{A truncated-normal distribution was defined for the priors of periods and masses of the planets as $\mathcal{TN}(\mu_0=\mu, \sigma^{2}_0=s\sigma$, lower\_limit=$\mu$-k$\sigma$, upper\_limit=$\mu$+k$\sigma$), where $\mu$ and $\sigma$ are the reported mean and the error of each parameter. The width of the distribution and the truncation limits were defined by setting (s,k)=(2.5, 3) and (s,k)=(1, 5) for the periods and masses, respectively. No negatives masses were allowed.  A half-normal distribution was used for the eccentricities h-$\mathcal{N}(\mu_0,\sigma^{2}_0)$. The time results of the mean TTV models (RMS and maximum amplitude) are also reported.
    }
\end{table*}

\section{Conclusion}\label{sec:conclusion}

The observations obtained during this \cheops{} campaign represent the first large series of measurement on the system since the initial \kepler{} data reported three-and-a-half years prior. The newly computed ephemerides for the system represent a leap forward in accuracy. Particularly for planets $d$, $f,$ and $g,$ the refined values are crucial for targeting future transit events. This is even more valuable as the accumulated deviations for these planets had already reached up to 200 minutes. For example, in the case of planet $f$, we examined the deviation between predictions by the end of 2024. Using the ephemerides from A22, we predict $T_C=2460673.05\pm0.12$. However, with the new average period derived from our work, we obtain $T_C=2460672.727\pm0.015$. The new ephemeris yield a $T_C$ 8 hours prior to that of A22.

We also want to emphasize the absence of transits of planet $b$ in \cheops{} light curves. Indeed, despite the numerous transit events predicted within our observing windows, no transits were confirmed in the \cheops{} dataset. 
The non-detection of the transits of planet $b$ in \cheops{} could result  from a large accumulated error in the ephemeris or, for example, from a dynamical mechanism such as the precession of the planet. Therefore, a dedicated photometric follow-up is required to resolve this debate.

Planets in the \system\ system typically display TTVs on the order of 10 minutes, except for planet $d,$ which shows TTV amplitudes of up to 60 minutes compared to the mean orbital period. However, \lang{this} is only possible if planet $d$ is of a mass of $\sim0.8$\mearth, with a fairly high eccentricity of $\sim0.16$.
As planet $d$ is in a MMR with its neigbors, large TTVs can be expected. Therefore, a future analysis based on better sampled TTVs will provide a more detailed view of the architecture of this compact system and its stability.
%This seems, however, unlikely due to the architecture of the system, and is most likely an artifact of the resonance chain.

%A future detailed analysis of the stability of this compact system could allow us to constrain this hypothesis further. In addition to continuous transit monitoring, a new high-precision radial velocity (RV) measurement campaign is essential to more accurately model TTVs. This campaign could be conducted using instruments such as \harps{}, as demonstrated in \citet{lopez2019}, or the ESPRESSO spectrograph.

%--------------------------------------------------------------------
% Acknowledgements
%--------------------------------------------------------------------

\begin{acknowledgements}
    HV and MD acknowledge funding from the Institut Universitaire de France (IUF) that made this work possible. SS acknowledges support from the Programme National de Planétologie (PNP) of CNRS-INSU. SH acknowledges CNES funding through the grant 837319. We thank the referee for their comments that improved the quality of the paper.
\end{acknowledgements}

\bibliography{biblio.bib}

\begin{thebibliography}{39}
\expandafter\ifx\csname natexlab\endcsname\relax\def\natexlab#1{#1}\fi

\bibitem[{{Acu{\~n}a} {et~al.}(2022){Acu{\~n}a}, {Lopez}, {Morel}, {Deleuil},
  {Mousis}, {Aguichine}, {Marcq}, \& {Santerne}}]{acuna2022}
{Acu{\~n}a}, L., {Lopez}, T.~A., {Morel}, T., {et~al.} 2022, \aap, 660, A102

\bibitem[{{Agol} \& {Deck}(2016)}]{AgolDeck2016}
{Agol}, E. \& {Deck}, K. 2016, \apj, 818, 177

\bibitem[{Agol {et~al.}(2005)Agol, Steffen, Sari, \& Clarkson}]{agol2005}
Agol, E., Steffen, J., Sari, R., \& Clarkson, W. 2005, Monthly Notices of the
  Royal Astronomical Society, 359, 567

\bibitem[{{Benz} {et~al.}(2021){Benz}, {Broeg}, {Fortier}, {Rando}, {Beck},
  {Beck}, {Queloz}, {Ehrenreich}, {Maxted}, {Isaak}, {Billot}, {Alibert},
  {Alonso}, {Ant{\'o}nio}, {Asquier}, {Bandy}, {B{\'a}rczy}, {Barrado},
  {Barros}, {Baumjohann}, {Bekkelien}, {Bergomi}, {Biondi}, {Bonfils},
  {Borsato}, {Brandeker}, {Busch}, {Cabrera}, {Cessa}, {Charnoz}, {Chazelas},
  {Collier Cameron}, {Corral Van Damme}, {Cortes}, {Davies}, {Deleuil},
  {Deline}, {Delrez}, {Demangeon}, {Demory}, {Erikson}, {Farinato}, {Fossati},
  {Fridlund}, {Futyan}, {Gandolfi}, {Garcia Munoz}, {Gillon}, {Guterman},
  {Gutierrez}, {Hasiba}, {Heng}, {Hernandez}, {Hoyer}, {Kiss}, {Kovacs},
  {Kuntzer}, {Laskar}, {Lecavelier des Etangs}, {Lendl}, {L{\'o}pez}, {Lora},
  {Lovis}, {L{\"u}ftinger}, {Magrin}, {Malvasio}, {Marafatto}, {Michaelis}, {de
  Miguel}, {Modrego}, {Munari}, {Nascimbeni}, {Olofsson}, {Ottacher},
  {Ottensamer}, {Pagano}, {Palacios}, {Pall{\'e}}, {Peter}, {Piazza}, {Piotto},
  {Pizarro}, {Pollaco}, {Ragazzoni}, {Ratti}, {Rauer}, {Ribas}, {Rieder},
  {Rohlfs}, {Safa}, {Salatti}, {Santos}, {Scandariato}, {S{\'e}gransan},
  {Simon}, {Smith}, {Sordet}, {Sousa}, {Steller}, {Szab{\'o}}, {Szoke},
  {Thomas}, {Tschentscher}, {Udry}, {Van Grootel}, {Viotto}, {Walter},
  {Walton}, {Wildi}, \& {Wolter}}]{Benz2021}
{Benz}, W., {Broeg}, C., {Fortier}, A., {et~al.} 2021, Experimental Astronomy,
  51, 109

\bibitem[{{Boyle} {et~al.}(2022){Boyle}, {Christiansen}, {Vissapragada}, \&
  {Hardegree-Ullman}}]{boyle2022}
{Boyle}, A., {Christiansen}, J.~L., {Vissapragada}, S., \& {Hardegree-Ullman},
  K.~K. 2022, Research Notes of the American Astronomical Society, 6, 71

\bibitem[{{Buchner}(2021)}]{Buchner2021b}
{Buchner}, J. 2021, The Journal of Open Source Software, 6, 3001

\bibitem[{{Buchner}(2023)}]{Buchner2021a}
{Buchner}, J. 2023, Statistics Surveys, 17, 169

\bibitem[{{Cabrera} {et~al.}(2014){Cabrera}, {Csizmadia}, {Lehmann}, {Dvorak},
  {Gandolfi}, {Rauer}, {Erikson}, {Dreyer}, {Eigm{\"u}ller}, \&
  {Hatzes}}]{Kepler-90_Cabrera}
{Cabrera}, J., {Csizmadia}, S., {Lehmann}, H., {et~al.} 2014, \apj, 781, 18

\bibitem[{{Christiansen} {et~al.}(2018){Christiansen}, {Crossfield},
  {Barentsen}, {Lintott}, {Barclay}, {Simmons}, {Petigura}, {Schlieder},
  {Dressing}, {Vanderburg}, {Allen}, {McMaster}, {Miller}, {Veldthuis},
  {Allen}, {Wolfenbarger}, {Cox}, {Zemiro}, {Howard}, {Livingston}, {Sinukoff},
  {Catron}, {Grey}, {Kusch}, {Terentev}, {Vales}, \&
  {Kristiansen}}]{christiansen2018}
{Christiansen}, J.~L., {Crossfield}, I. J.~M., {Barentsen}, G., {et~al.} 2018,
  \aj, 155, 57

\bibitem[{{Dai} {et~al.}(2023){Dai}, {Masuda}, {Beard}, {Robertson},
  {Goldberg}, {Batygin}, {Bouma}, {Lissauer}, {Knudstrup}, {Albrecht},
  {Howard}, {Knutson}, {Petigura}, {Weiss}, {Isaacson}, {Kristiansen},
  {Osborn}, {Wang}, {Wang}, {Behmard}, {Greklek-McKeon}, {Vissapragada},
  {Batalha}, {Brinkman}, {Chontos}, {Crossfield}, {Dressing}, {Fetherolf},
  {Fulton}, {Hill}, {Huber}, {Kane}, {Lubin}, {MacDougall}, {Mayo},
  {Mo{\v{c}}nik}, {Akana Murphy}, {Rubenzahl}, {Scarsdale}, {Tyler}, {Zandt},
  {Polanski}, {Schwengeler}, {Terentev}, {Benni}, {Bieryla}, {Ciardi}, {Falk},
  {Furlan}, {Girardin}, {Guerra}, {Hesse}, {Howell}, {Lillo-Box}, {Matthews},
  {Twicken}, {Villase{\~n}or}, {Latham}, {Jenkins}, {Ricker}, {Seager},
  {Vanderspek}, \& {Winn}}]{TOI-1136_Dai}
{Dai}, F., {Masuda}, K., {Beard}, C., {et~al.} 2023, \aj, 165, 33

\bibitem[{{Delrez} {et~al.}(2023){Delrez}, {Leleu}, {Brandeker}, {Gillon},
  {Hooton}, {Collier Cameron}, {Deline}, {Fortier}, {Queloz}, {Bonfanti}, {Van
  Grootel}, {Wilson}, {Egger}, {Alibert}, {Alonso}, {Anglada}, {Asquier},
  {B{\'a}rczy}, {Barrado y Navascues}, {Barros}, {Baumjohann}, {Beck}, {Beck},
  {Benz}, {Billot}, {Bonf{\i}ls}, {Borsato}, {Broeg}, {Buder}, {Cabrera},
  {Cessa}, {Charnoz}, {Csizmadia}, {Cubillos}, {Davies}, {Deleuil},
  {Demangeon}, {Demory}, {Ehrenreich}, {Erikson}, {Fossati}, {Fridlund},
  {Gandolfi}, {G{\"u}del}, {Hasiba}, {Hoyer}, {Isaak}, {Jenkins}, {Kiss},
  {Laskar}, {Latham}, {Lecavelier des Etangs}, {Lendl}, {Lovis}, {Luque},
  {Magrin}, {Maxted}, {Mordasini}, {Nascimbeni}, {Olofsson}, {Ottensamer},
  {Pagano}, {Pall{\'e}}, {Peter}, {Piotto}, {Pollacco}, {Ragazzoni}, {Rando},
  {Rauer}, {Ribas}, {Ricker}, {Santos}, {Scandariato}, {Seager},
  {S{\'e}gransan}, {Simon}, {Smith}, {Sousa}, {Steller}, {Szab{\'o}}, {Thomas},
  {Udry}, {Vanderspek}, {Venturini}, {Viotto}, {Walton}, \&
  {Winn}}]{TOI-178-Delrez}
{Delrez}, L., {Leleu}, A., {Brandeker}, A., {et~al.} 2023, \aap, 678, A200

\bibitem[{{Espinoza} {et~al.}(2019){Espinoza}, {Kossakowski}, \&
  {Brahm}}]{juliet}
{Espinoza}, N., {Kossakowski}, D., \& {Brahm}, R. 2019, \mnras, 490, 2262

\bibitem[{Fabrycky {et~al.}(2014)Fabrycky, Lissauer, Ragozzine, Rowe, Steffen,
  Agol, Barclay, Batalha, Borucki, Ciardi, Ford, Gautier, Geary, Holman,
  Jenkins, Li, Morehead, Morris, Shporer, Smith, Still, \&
  Van~Cleve}]{fabrycky2014}
Fabrycky, D.~C., Lissauer, J.~J., Ragozzine, D., {et~al.} 2014, The
  Astrophysical Journal, 790, 146

\bibitem[{{Foreman-Mackey} {et~al.}(2013){Foreman-Mackey}, {Hogg}, {Lang}, \&
  {Goodman}}]{EMCEE}
{Foreman-Mackey}, D., {Hogg}, D.~W., {Lang}, D., \& {Goodman}, J. 2013, \pasp,
  125, 306

\bibitem[{{Gaia Collaboration} {et~al.}(2018){Gaia Collaboration}, {Brown},
  {Vallenari}, {Prusti}, {de Bruijne}, {Babusiaux}, {Bailer-Jones}, {Biermann},
  {Evans}, {Eyer}, \& et~al.}]{gaia_dr2}
{Gaia Collaboration}, {Brown}, A.~G.~A., {Vallenari}, A., {et~al.} 2018, \aap,
  616, A1

\bibitem[{{Gillon} {et~al.}(2017){Gillon}, {Triaud}, {Demory}, {Jehin}, {Agol},
  {Deck}, {Lederer}, {de Wit}, {Burdanov}, {Ingalls}, {Bolmont}, {Leconte},
  {Raymond}, {Selsis}, {Turbet}, {Barkaoui}, {Burgasser}, {Burleigh}, {Carey},
  {Chaushev}, {Copperwheat}, {Delrez}, {Fernandes}, {Holdsworth}, {Kotze}, {Van
  Grootel}, {Almleaky}, {Benkhaldoun}, {Magain}, \&
  {Queloz}}]{TRAPPIST-1-refined}
{Gillon}, M., {Triaud}, A. H.~M.~J., {Demory}, B.-O., {et~al.} 2017, \nat, 542,
  456

\bibitem[{{Goldberg} \& {Batygin}(2022)}]{Goldberg2022}
{Goldberg}, M. \& {Batygin}, K. 2022, \aj, 163, 201

\bibitem[{{Hara} {et~al.}(2020){Hara}, {Bouchy}, {Stalport}, {Boisse},
  {Rodrigues}, {Delisle}, {Santerne}, {Henry}, {Arnold}, {Astudillo-Defru},
  {Borgniet}, {Bonfils}, {Bourrier}, {Brugger}, {Courcol}, {Dalal}, {Deleuil},
  {Delfosse}, {Demangeon}, {D{\'\i}az}, {Dumusque}, {Forveille}, {H{\'e}brard},
  {Hobson}, {Kiefer}, {Lopez}, {Mignon}, {Mousis}, {Moutou}, {Pepe}, {Rey},
  {Santos}, {S{\'e}gransan}, {Udry}, \& {Wilson}}]{HD158259}
{Hara}, N.~C., {Bouchy}, F., {Stalport}, M., {et~al.} 2020, \aap, 636, L6

\bibitem[{{Hardegree-Ullman} {et~al.}(2021){Hardegree-Ullman}, {Christiansen},
  {Ciardi}, {Crossfield}, {Dressing}, {Livingston}, {Volk}, {Agol}, {Barclay},
  {Barentsen}, {Benneke}, {Gorjian}, \& {Kristiansen}}]{hardegree-ullman2021}
{Hardegree-Ullman}, K.~K., {Christiansen}, J.~L., {Ciardi}, D.~R., {et~al.}
  2021, \aj, 161, 219

\bibitem[{Holman \& Murray(2005)}]{holman2005}
Holman, M.~J. \& Murray, N.~W. 2005, Science, 307, 1288

\bibitem[{{Howell} {et~al.}(2014){Howell}, {Sobeck}, {Haas}, {Still},
  {Barclay}, {Mullally}, {Troeltzsch}, {Aigrain}, {Bryson}, {Caldwell},
  {Chaplin}, {Cochran}, {Huber}, {Marcy}, {Miglio}, {Najita}, {Smith},
  {Twicken}, \& {Fortney}}]{Howell2014}
{Howell}, S.~B., {Sobeck}, C., {Haas}, M., {et~al.} 2014, \pasp, 126, 398

\bibitem[{{Hoyer} {et~al.}(2022){Hoyer}, {Bonfanti}, {Leleu}, {Acu{\~n}a},
  {Serrano}, {Deleuil}, {Bekkelien}, {Broeg}, {Flor{\'e}n}, {Queloz}, {Wilson},
  {Sousa}, {Hooton}, {Adibekyan}, {Alibert}, {Alonso}, {Anglada}, {Asquier},
  {B{\'a}rczy}, {Barrado}, {Barros}, {Baumjohann}, {Beck}, {Beck}, {Benz},
  {Billot}, {Biondi}, {Bonfils}, {Brandeker}, {Cabrera}, {Charnoz}, {Collier
  Cameron}, {Csizmadia}, {Davies}, {Delrez}, {Demangeon}, {Demory},
  {Ehrenreich}, {Erikson}, {Fortier}, {Fossati}, {Fridlund}, {Gandolfi},
  {Gillon}, {G{\"u}del}, {Hara}, {Heng}, {Isaak}, {Jenkins}, {Kiss}, {Laskar},
  {Latham}, {Lecavelier des Etangs}, {Lendl}, {Lovis}, {Luntzer}, {Magrin},
  {Maxted}, {Nascimbeni}, {Olofsson}, {Ottensamer}, {Pagano}, {Pall{\'e}},
  {Persson}, {Peter}, {Piazza}, {Piotto}, {Pollacco}, {Ragazzoni}, {Rando},
  {Rauer}, {Ribas}, {Ricker}, {Salmon}, {Santos}, {Scandariato}, {Seager},
  {S{\'e}gransan}, {Simon}, {Smith}, {Steller}, {Szab{\'o}}, {Thomas},
  {Twicken}, {Udry}, {Van Grootel}, {Vanderspek}, {Walton}, {Westerdorff}, \&
  {Winn}}]{HD108236_Hoyer2022}
{Hoyer}, S., {Bonfanti}, A., {Leleu}, A., {et~al.} 2022, \aap, 668, A117

\bibitem[{{Hoyer} {et~al.}(2020){Hoyer}, {Guterman}, {Demangeon}, {Sousa},
  {Deleuil}, {Meunier}, \& {Benz}}]{hoyer2020}
{Hoyer}, S., {Guterman}, P., {Demangeon}, O., {et~al.} 2020, \aap, 635, A24

\bibitem[{{Hoyer} {et~al.}(2023){Hoyer}, {Jenkins}, {Parmentier}, {Deleuil},
  {Scandariato}, {Wilson}, {D{\'\i}az}, {Crossfield}, {Dragomir}, {Kataria},
  {Lendl}, {Ramirez}, {Pe{\~n}a Rojas}, \& {Vin{\'e}s}}]{Hoyer_2023}
{Hoyer}, S., {Jenkins}, J.~S., {Parmentier}, V., {et~al.} 2023, \aap, 675, A81

\bibitem[{{Leleu} {et~al.}(2021){Leleu}, {Alibert}, {Hara}, {Hooton}, {Wilson},
  {Robutel}, {Delisle}, {Laskar}, {Hoyer}, {Lovis}, {Bryant}, {Ducrot},
  {Cabrera}, {Delrez}, {Acton}, {Adibekyan}, {Allart}, {Allende Prieto},
  {Alonso}, {Alves}, {Anderson}, {Angerhausen}, {Anglada Escud{\'e}},
  {Asquier}, {Barrado}, {Barros}, {Baumjohann}, {Bayliss}, {Beck}, {Beck},
  {Bekkelien}, {Benz}, {Billot}, {Bonfanti}, {Bonfils}, {Bouchy}, {Bourrier},
  {Bou{\'e}}, {Brandeker}, {Broeg}, {Buder}, {Burdanov}, {Burleigh},
  {B{\'a}rczy}, {Cameron}, {Chamberlain}, {Charnoz}, {Cooke}, {Corral Van
  Damme}, {Correia}, {Cristiani}, {Damasso}, {Davies}, {Deleuil}, {Demangeon},
  {Demory}, {Di Marcantonio}, {Di Persio}, {Dumusque}, {Ehrenreich}, {Erikson},
  {Figueira}, {Fortier}, {Fossati}, {Fridlund}, {Futyan}, {Gandolfi},
  {Garc{\'\i}a Mu{\~n}oz}, {Garcia}, {Gill}, {Gillen}, {Gillon}, {Goad},
  {Gonz{\'a}lez Hern{\'a}ndez}, {Guedel}, {G{\"u}nther}, {Haldemann},
  {Henderson}, {Heng}, {Hogan}, {Isaak}, {Jehin}, {Jenkins}, {Jord{\'a}n},
  {Kiss}, {Kristiansen}, {Lam}, {Lavie}, {Lecavelier des Etangs}, {Lendl},
  {Lillo-Box}, {Lo Curto}, {Magrin}, {Martins}, {Maxted}, {McCormac}, {Mehner},
  {Micela}, {Molaro}, {Moyano}, {Murray}, {Nascimbeni}, {Nunes}, {Olofsson},
  {Osborn}, {Oshagh}, {Ottensamer}, {Pagano}, {Pall{\'e}}, {Pedersen}, {Pepe},
  {Persson}, {Peter}, {Piotto}, {Polenta}, {Pollacco}, {Poretti}, {Pozuelos},
  {Queloz}, {Ragazzoni}, {Rando}, {Ratti}, {Rauer}, {Raynard}, {Rebolo},
  {Reimers}, {Ribas}, {Santos}, {Scandariato}, {Schneider}, {Sebastian},
  {Sestovic}, {Simon}, {Smith}, {Sousa}, {Sozzetti}, {Steller}, {Su{\'a}rez
  Mascare{\~n}o}, {Szab{\'o}}, {S{\'e}gransan}, {Thomas}, {Thompson},
  {Tilbrook}, {Triaud}, {Turner}, {Udry}, {Van Grootel}, {Venus}, {Verrecchia},
  {Vines}, {Walton}, {West}, {Wheatley}, {Wolter}, \& {Zapatero
  Osorio}}]{TOI-178-Leleu}
{Leleu}, A., {Alibert}, Y., {Hara}, N.~C., {et~al.} 2021, \aap, 649, A26

\bibitem[{{Lissauer} {et~al.}(2014){Lissauer}, {Marcy}, {Bryson}, {Rowe},
  {Jontof-Hutter}, {Agol}, {Borucki}, {Carter}, {Ford}, {Gilliland}, {Kolbl},
  {Star}, {Steffen}, \& {Torres}}]{Lissauer_2014}
{Lissauer}, J.~J., {Marcy}, G.~W., {Bryson}, S.~T., {et~al.} 2014, \apj, 784,
  44

\bibitem[{Lissauer {et~al.}(2011)Lissauer, Ragozzine, Fabrycky, Steffen, Ford,
  Jenkins, Shporer, Holman, Rowe, Quintana, Batalha, Borucki, Bryson, Caldwell,
  Carter, Ciardi, Dunham, Fortney, Gautier, Howell, Koch, Latham, Marcy,
  Morehead, \& Sasselov}]{lissauer2011}
Lissauer, J.~J., Ragozzine, D., Fabrycky, D.~C., {et~al.} 2011, The
  Astrophysical Journal Supplement Series, 197, 8

\bibitem[{{Lopez} {et~al.}(2019){Lopez}, {Barros}, {Santerne}, {Deleuil},
  {Adibekyan}, {Almenara}, {Armstrong}, {Brugger}, {Barrado}, {Bayliss},
  {Boisse}, {Bonomo}, {Bouchy}, {Brown}, {Carli}, {Demangeon}, {Dumusque},
  {D{\'\i}az}, {Faria}, {Figueira}, {Foxell}, {Giles}, {H{\'e}brard},
  {Hojjatpanah}, {Kirk}, {Lillo-Box}, {Lovis}, {Mousis}, {da N{\'o}brega},
  {Nielsen}, {Neal}, {Osborn}, {Pepe}, {Pollacco}, {Santos}, {Sousa}, {Udry},
  {Vigan}, \& {Wheatley}}]{lopez2019}
{Lopez}, T.~A., {Barros}, S.~C.~C., {Santerne}, A., {et~al.} 2019, \aap, 631,
  A90

\bibitem[{{Luger} {et~al.}(2017){Luger}, {Sestovic}, {Kruse}, {Grimm},
  {Demory}, {Agol}, {Bolmont}, {Fabrycky}, {Fernandes}, {Van Grootel},
  {Burgasser}, {Gillon}, {Ingalls}, {Jehin}, {Raymond}, {Selsis}, {Triaud},
  {Barclay}, {Barentsen}, {Howell}, {Delrez}, {de Wit}, {Foreman-Mackey},
  {Holdsworth}, {Leconte}, {Lederer}, {Turbet}, {Almleaky}, {Benkhaldoun},
  {Magain}, {Morris}, {Heng}, \& {Queloz}}]{TRAPPIST-1_Luger}
{Luger}, R., {Sestovic}, M., {Kruse}, E., {et~al.} 2017, Nature Astronomy, 1,
  0129

\bibitem[{{Luque} {et~al.}(2023){Luque}, {Osborn}, {Leleu}, {Pall{\'e}},
  {Bonfanti}, {Barrag{\'a}n}, {Wilson}, {Broeg}, {Cameron}, {Lendl}, {Maxted},
  {Alibert}, {Gandolfi}, {Delisle}, {Hooton}, {Egger}, {Nowak}, {Lafarga},
  {Rapetti}, {Twicken}, {Morales}, {Carleo}, {Orell-Miquel}, {Adibekyan},
  {Alonso}, {Alqasim}, {Amado}, {Anderson}, {Anglada-Escud{\'e}}, {Bandy},
  {B{\'a}rczy}, {Barrado Navascues}, {Barros}, {Baumjohann}, {Bayliss}, {Bean},
  {Beck}, {Beck}, {Benz}, {Billot}, {Bonfils}, {Borsato}, {Boyle}, {Brandeker},
  {Bryant}, {Cabrera}, {Carrazco-Gaxiola}, {Charbonneau}, {Charnoz}, {Ciardi},
  {Cochran}, {Collins}, {Crossfield}, {Csizmadia}, {Cubillos}, {Dai}, {Davies},
  {Deeg}, {Deleuil}, {Deline}, {Delrez}, {Demangeon}, {Demory}, {Ehrenreich},
  {Erikson}, {Esparza-Borges}, {Falk}, {Fortier}, {Fossati}, {Fridlund},
  {Fukui}, {Garcia-Mejia}, {Gill}, {Gillon}, {Goffo}, {G{\'o}mez Maqueo Chew},
  {G{\"u}del}, {Guenther}, {G{\"u}nther}, {Hatzes}, {Helling}, {Hesse},
  {Howell}, {Hoyer}, {Ikuta}, {Isaak}, {Jenkins}, {Kagetani}, {Kiss}, {Kodama},
  {Korth}, {Lam}, {Laskar}, {Latham}, {Lecavelier des Etangs}, {Leon},
  {Livingston}, {Magrin}, {Matson}, {Matthews}, {Mordasini}, {Mori}, {Moyano},
  {Munari}, {Murgas}, {Narita}, {Nascimbeni}, {Olofsson}, {Osborne},
  {Ottensamer}, {Pagano}, {Parviainen}, {Peter}, {Piotto}, {Pollacco},
  {Queloz}, {Quinn}, {Quirrenbach}, {Ragazzoni}, {Rando}, {Ratti}, {Rauer},
  {Redfield}, {Ribas}, {Ricker}, {Rudat}, {Sabin}, {Salmon}, {Santos},
  {Scandariato}, {Schanche}, {Schlieder}, {Seager}, {S{\'e}gransan}, {Shporer},
  {Simon}, {Smith}, {Sousa}, {Stalport}, {Szab{\'o}}, {Thomas}, {Tuson},
  {Udry}, {Vanderburg}, {Van Eylen}, {Van Grootel}, {Venturini}, {Walter},
  {Walton}, {Watanabe}, {Winn}, \& {Zingales}}]{Luque2023}
{Luque}, R., {Osborn}, H.~P., {Leleu}, A., {et~al.} 2023, \nat, 623, 932

\bibitem[{Luque {et~al.}(2023)Luque, Osborn, Leleu, Pall{\'e}, Bonfanti,
  Barrag{\'a}n, Wilson, Broeg, Cameron, Lendl, Maxted, Alibert, Gandolfi,
  Delisle, Hooton, Egger, Nowak, Lafarga, Rapetti, Twicken, Morales, Carleo,
  Orell-Miquel, Adibekyan, Alonso, Alqasim, Amado, Anderson,
  Anglada-Escud{\'e}, Bandy, B{\'a}rczy, Barrado~Navascues, Barros, Baumjohann,
  Bayliss, Bean, Beck, Beck, Benz, Billot, Bonfils, Borsato, Boyle, Brandeker,
  Bryant, Cabrera, Carrazco-Gaxiola, Charbonneau, Charnoz, Ciardi, Cochran,
  Collins, Crossfield, Csizmadia, Cubillos, Dai, Davies, Deeg, Deleuil, Deline,
  Delrez, Demangeon, Demory, Ehrenreich, Erikson, Esparza-Borges, Falk,
  Fortier, Fossati, Fridlund, Fukui, Garcia-Mejia, Gill, Gillon, Goffo,
  G{\'o}mez Maqueo~Chew, G{\"u}del, Guenther, G{\"u}nther, Hatzes, Helling,
  Hesse, Howell, Hoyer, Ikuta, Isaak, Jenkins, Kagetani, Kiss, Kodama, Korth,
  Lam, Laskar, Latham, Lecavelier~des Etangs, Leon, Livingston, Magrin, Matson,
  Matthews, Mordasini, Mori, Moyano, Munari, Murgas, Narita, Nascimbeni,
  Olofsson, Osborne, Ottensamer, Pagano, Parviainen, Peter, Piotto, Pollacco,
  Queloz, Quinn, Quirrenbach, Ragazzoni, Rando, Ratti, Rauer, Redfield, Ribas,
  Ricker, Rudat, Sabin, Salmon, Santos, Scandariato, Schanche, Schlieder,
  Seager, S{\'e}gransan, Shporer, Simon, Smith, Sousa, Stalport, Szab{\'o},
  Thomas, Tuson, Udry, Vanderburg, Van~Eylen, Van~Grootel, Venturini, Walter,
  Walton, Watanabe, Winn, \& Zingales}]{HD110067}
Luque, R., Osborn, H.~P., Leleu, A., {et~al.} 2023, Nature, 623, 932

\bibitem[{{Maxted} {et~al.}(2022){Maxted}, {Ehrenreich}, {Wilson}, {Alibert},
  {Cameron}, {Hoyer}, {Sousa}, {Olofsson}, {Bekkelien}, {Deline}, {Delrez},
  {Bonfanti}, {Borsato}, {Alonso}, {Anglada Escud{\'e}}, {Barrado}, {Barros},
  {Baumjohann}, {Beck}, {Beck}, {Benz}, {Billot}, {Biondi}, {Bonfils},
  {Brandeker}, {Broeg}, {B{\'a}rczy}, {Cabrera}, {Charnoz}, {Corral Van Damme},
  {Csizmadia}, {Davies}, {Deleuil}, {Demangeon}, {Demory}, {Erikson},
  {Flor{\'e}n}, {Fortier}, {Fossati}, {Fridlund}, {Futyan}, {Gandolfi},
  {Gillon}, {Guedel}, {Guterman}, {Heng}, {Isaak}, {Kiss}, {Laskar},
  {Lecavelier des Etangs}, {Lendl}, {Lovis}, {Magrin}, {Nascimbeni},
  {Ottensamer}, {Pagano}, {Pall{\'e}}, {Peter}, {Piotto}, {Pollacco},
  {Pozuelos}, {Queloz}, {Ragazzoni}, {Rando}, {Rauer}, {Reimers}, {Ribas},
  {Salmon}, {Santos}, {Scandariato}, {Simon}, {Smith}, {Steller}, {Swayne},
  {Szab{\'o}}, {S{\'e}gransan}, {Thomas}, {Udry}, {Van Grootel}, \&
  {Walton}}]{pycheops}
{Maxted}, P.~F.~L., {Ehrenreich}, D., {Wilson}, T.~G., {et~al.} 2022, \mnras,
  514, 77

\bibitem[{Miralda-Escude(2002)}]{miralda-escude2002}
Miralda-Escude, J. 2002, The Astrophysical Journal, 564, 1019,
  arXiv:astro-ph/0104034

\bibitem[{{Mishra} {et~al.}(2021){Mishra}, {Alibert}, {Leleu}, {Emsenhuber},
  {Mordasini}, {Burn}, {Udry}, \& {Benz}}]{Mishra_2021}
{Mishra}, L., {Alibert}, Y., {Leleu}, A., {et~al.} 2021, \aap, 656, A74

\bibitem[{{Mishra} {et~al.}(2023){Mishra}, {Alibert}, {Udry}, \&
  {Mordasini}}]{mishra2023}
{Mishra}, L., {Alibert}, Y., {Udry}, S., \& {Mordasini}, C. 2023, \aap, 670,
  A68

\bibitem[{{Newville} {et~al.}(2014){Newville}, {Stensitzki}, {Allen}, \&
  {Ingargiola}}]{LMFIT}
{Newville}, M., {Stensitzki}, T., {Allen}, D.~B., \& {Ingargiola}, A. 2014,
  {LMFIT: Non-Linear Least-Square Minimization and Curve-Fitting for Python},
  Zenodo

\bibitem[{{Shallue} \& {Vanderburg}(2018)}]{Shallue_2018}
{Shallue}, C.~J. \& {Vanderburg}, A. 2018, \aj, 155, 94

\bibitem[{{Vogt} {et~al.}(2015){Vogt}, {Burt}, {Meschiari}, {Butler}, {Henry},
  {Wang}, {Holden}, {Gapp}, {Hanson}, {Arriagada}, {Keiser}, {Teske}, \&
  {Laughlin}}]{HD219134}
{Vogt}, S.~S., {Burt}, J., {Meschiari}, S., {et~al.} 2015, \apj, 814, 12

\bibitem[{{Xie}(2013)}]{Xie2013}
{Xie}, J.-W. 2013, \apjs, 208, 22

\end{thebibliography}
\bibliographystyle{aa}

\begin{appendix}

\section{System parameters}\label{app:parameters}

This appendix compiles the stellar and planetary parameters used in this study. Table~\ref{table:stellar_parameters} shows the stellar parameters of host star and Table~\ref{table:planetary_parameters} lists the parameters of the planets in the system.

\renewcommand{\arraystretch}{1}

\renewcommand{\arraystretch}{1.3} % Better spacing for readability
\begin{table*}[]

\caption{Priors used for the modeling of the \cheops{} multi-visit, and the resulting best \texttt{pycheops} fit.}
    \label{table:planetary_parameters}
    \centering
    \begin{tabular}{ l c c }
        \hline
        \textbf{Parameters} & Prior & Best \cheops{} multi-visit fit \\
        \hline\hline
        
        \multicolumn{3}{c}{Planet d} \\
        \hline
        Period [days] & $5.40479 \pm 0.00021$ & $5.40533 \pm 0.00043$ \\
        $T_0$ [BJD-2450000] & $7743.15984^{0.00095}_{-0.00093}$ & $9110.7186 \pm 0.0024$ \\
        Depth [ppm] & $645.16^{+35.05}_{-33.02}$ & $837.98 \pm 117.64$ \\
        Transit duration $T_{14}$ [h] & $2.71^{+0.07}_{-0.08}$ & $2.758 \pm 0.094$ \\
        Impact parameter $b$ & $0.297^{+0.145}_{-0.170}$ & $0.388 \pm 0.215$ \\
        \hline
        
        \multicolumn{3}{c}{Planet e} \\
        \hline
        Period [days] & $8.26146^{+0.00022}_{-0.00021}$ & $8.26129 \pm 0.00042$ \\
        $T_0$ [BJD-2450000] & $7740.64563^{+0.00085}_{-0.00087}$ & $9095.5044 \pm 0.0021$ \\
        Depth [ppm] & $1298.88^{+55.34}_{-51.90}$ & $1285.08 \pm 96.49$ \\
        Transit duration $T_{14}$ [h] & $2.97 \pm 0.05$ & $2.994 \pm 0.0625$ \\
        Impact parameter $b$ & $0.474^{+0.081}_{-0.115}$ & $0.551 \pm 0.111$ \\
        \hline
        
        \multicolumn{3}{c}{Planet f} \\
        \hline
        Period [days] & $12.75760^{+0.00051}_{-0.00048}$ & $12.7581 \pm 0.00137$ \\
        $T_0$ [BJD-2450000] & $7738.80226^{+0.00093}_{-0.00092}$ & $9090.9061 \pm 0.00571$ \\
        Depth [ppm] & $939.42^{+52.10}_{-50.88}$ & $850.69 \pm 126.55$ \\
        Transit duration $T_{14}$ [h] & $3.2 \pm 0.08$ & $3.240 \pm 0.112$ \\
        Impact parameter $b$ & $0.541^{+0.073}_{-0.109}$ & $0.525 \pm 0.145$ \\
        \hline
        
        \multicolumn{3}{c}{Planet g} \\
        \hline
        Period [days] & $41.96822^{+0.00817}_{-0.00774}$ & $41.96851 \pm 0.00786$ \\
        $T_0$ [BJD-2450000] & $7773.86006^{+0.01931}_{-0.03522}$ & $9116.7492 \pm 0.0020$ \\
        Depth [ppm] & $1023.36^{+209.21}_{-158.67}$ & $841.49 \pm 189.47$ \\
        Transit duration $T_{14}$ [h] & $4.71^{+0.79}_{-1.63}$ & $4.226 \pm 0.027$ \\
        Impact parameter $b$ & $0.55^{+0.319}_{-0.365}$ & $0.487 \pm 0.231$ \\
        \hline
    \end{tabular}
    \tablefoot{As planet $g$ only has a single visit, the resulting fit values are therefore from the single visit analysis. These results are solely based on the transits acquired by \cheops, the ephemerides from the complete analysis can be found in Table~\ref{tab:new_ephemeris}.}
    
\end{table*}
\renewcommand{\arraystretch}{1.2}

\section{Multi-transit figures}\label{app:multi}

Here, we present all the light curves used during multi-transit analyses. \lang{All the}  figures show the best-fit models obtained with \texttt{pycheops}. For each figure the detrented data are shown as gray dots, with their 15 minute bins as black circles. The transit model is shown as a black-gray line. The transit are plotted in chronological order from bottom to top.

\begin{figure}[h]
    \centering
    \includegraphics[width=\linewidth]{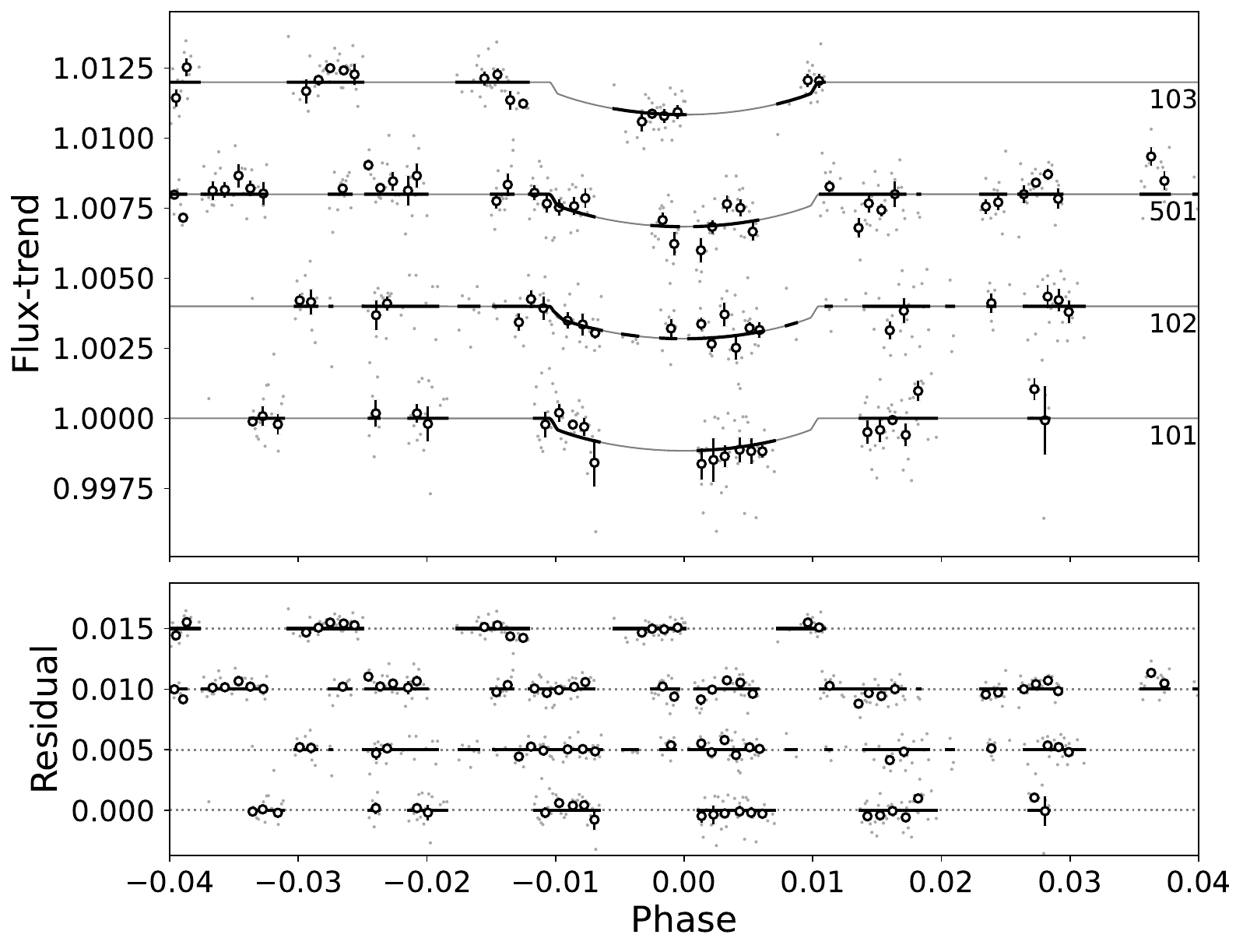}
    \caption{Transits and the best-fit model for the multi-visit of planet $d$.}
    \label{fig:d-transit}
\end{figure}

\begin{figure}[h]
    \centering
    \includegraphics[width=\linewidth]{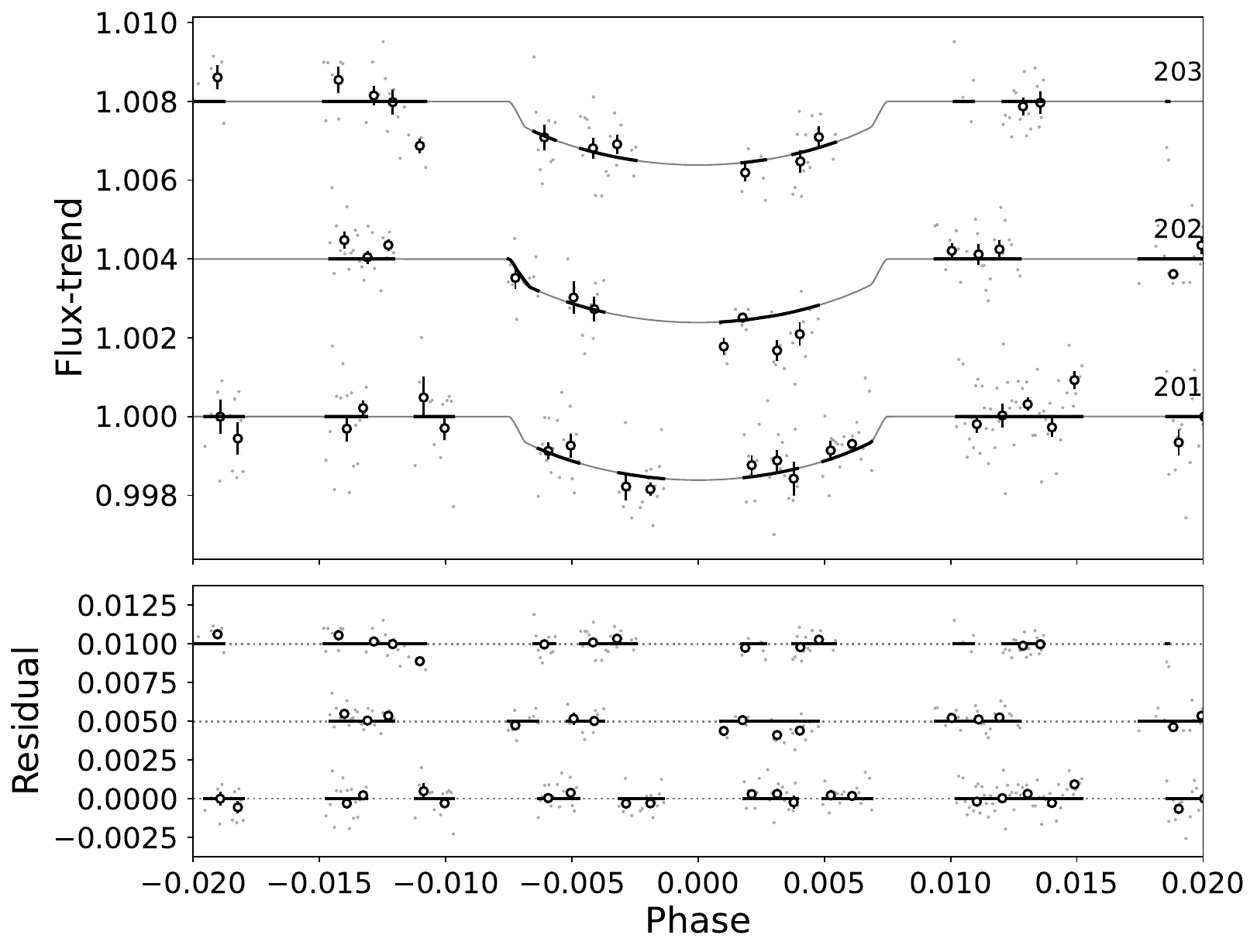}
    \caption{Transits and the best-fit model for the multi-visit of planet $e$.}
    \label{fig:e-transit}
\end{figure}

\begin{figure}[h]
    \centering
    \includegraphics[width=\linewidth]{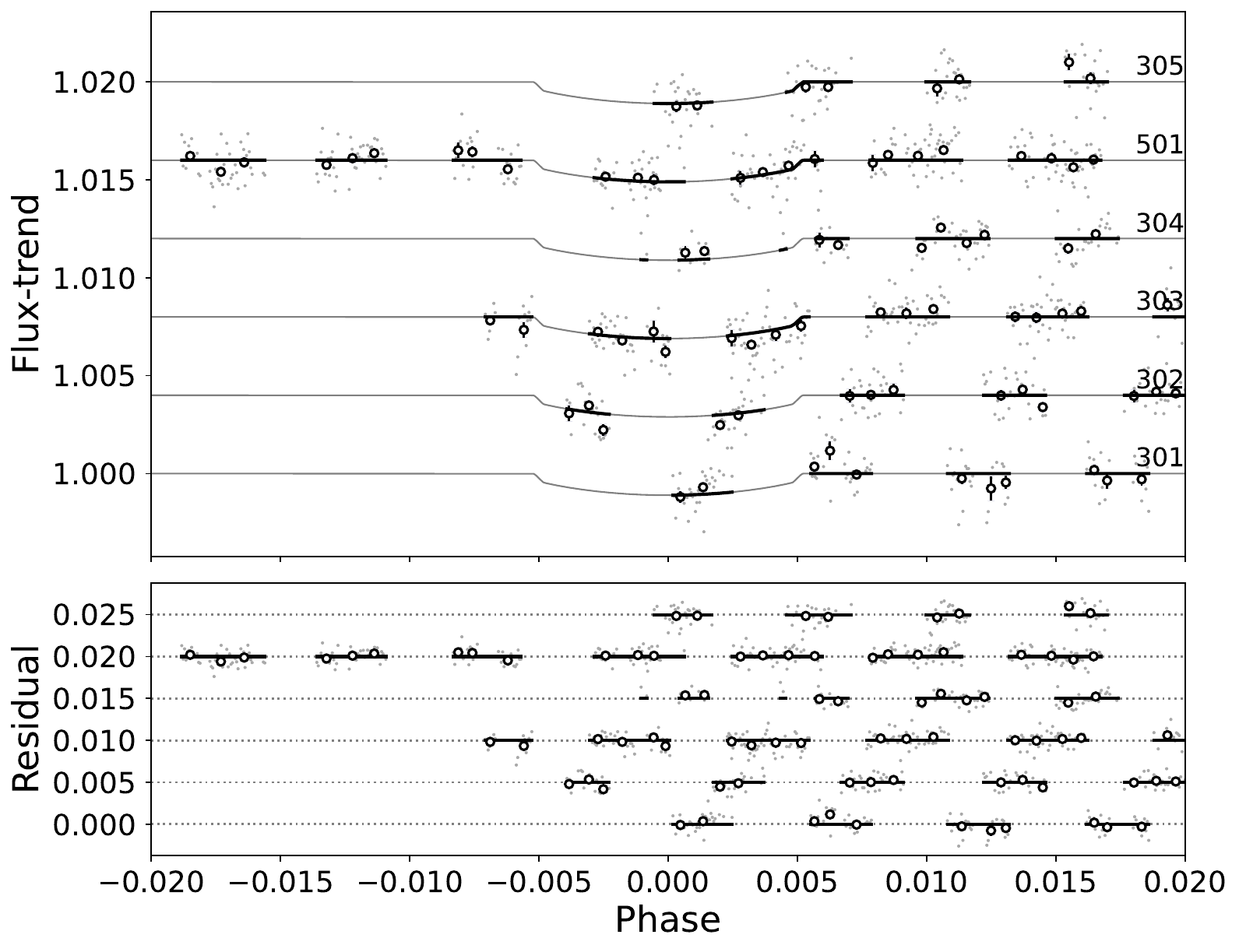}
    \caption{Transits and the \lang{best-fit} model for the multi-visit of planet $f$.}
    \label{fig:f-transit}
\end{figure}

\section{O-C diagram}\label{app:oc}

\begin{table*}[]

 \caption{Transit times and deviations of the $b$, $c$, and $d$ planets in the $K$2-138 system.}
    \label{tab:ttvs}
    \centering
    \begin{tabular}{ c c c c l}
    
\hline
\hline
Planet & Epoch & $T_\mathrm{C}-$2~450~000 & O--C & Source \\
       &       & [BJD\_TDB] & [min] & \\
\hline
$b$
& 0 & 7740.3713 $\pm$ 0.0060 & -- & HU21 \\
& 14 & 7773.3168 $\pm$ 0.0009 & -- & L19 \\
& 14 & 7773.3170 $\pm$ 0.0038 & -- & Ch18 \\
\hline

$c$
&0  & 7740.3218 $\pm$ 0.0009 & 0 $\pm$ 1 & L19 (T$_0$)\\
&0  & 7740.3223 $\pm$ 0.0025 & 1 $\pm$ 4 & Ch18 (T$_0$) \\
&0  & 7740.3210 $\pm$ 0.0030 & $-1$ $\pm$ 4 & HU21 (T$_0$)\\
&1      &7743.8862 $\pm$        0.0206 &        7 $\pm$ 30 & \kt\\
&2      &7747.4402 $\pm$        0.0057 &        $-2$ $\pm$ 8 & \kt\\
&3      &7751.0001 $\pm$        0.0089 &        $-2$ $\pm$ 13 & \kt\\
&4      &7754.5610 $\pm$        0.0113 &        0 $\pm$ 16 & \kt\\
&5      &7758.1188 $\pm$        0.0039 &        $-3$ $\pm$ 6 & \kt\\
&6      &7761.6797 $\pm$        0.0064 &        $-1$ $\pm$ 9 & \kt\\
&8      &7768.8083 $\pm$        0.0248 &         12 $\pm$ 36 & \kt\\
&10     &7775.9192 $\pm$        0.0054 &        $-1$ $\pm$ 8 & \kt\\
&11     &7779.4805 $\pm$        0.0056 &        1 $\pm$ 8 & \kt\\
&12     &7783.0410 $\pm$        0.0028 &        2 $\pm$ 4 & \kt\\
&13     &7786.5934 $\pm$        0.0089 &        $-9$ $\pm$ 13 & \kt\\
&15     &7793.7201 $\pm$        0.0079 &        2 $\pm$ 11 & \kt\\
&16     &7797.2763 $\pm$        0.0249 &  $-4$ $\pm$ 36 & \kt\\
&17     &7800.8394 $\pm$        0.0067 &        1 $\pm$ 10 & \kt\\
&18     &7804.4000 $\pm$        0.0067 &        2 $\pm$ 10 & \kt\\
&19     &7807.9484 $\pm$        0.0078 &        $-14$ $\pm$ 11 & \kt\\
&20     &7811.5195 $\pm$        0.0042 &        2 $\pm$ 6 & \kt\\
&21     &7815.0791 $\pm$        0.0092 &        2 $\pm$ 13 & \kt\\
&497    &9509.6957 $\pm$        0.0020 &        204 $\pm$ 36 & \cheops\\
\hline
$d$ &   0       &7743.1607 $\pm$        0.0041 & 7 $\pm$ 14 & \kt\\
&       1       &7748.5660 $\pm$        0.0043 &        7 $\pm$ 15 & \kt \\
&       2       &7753.9560 $\pm$        0.0242 &         $-15$ $\pm$ 37 & \kt\\
&       3       &7759.3751 $\pm$        0.0047 &        4 $\pm$ 15 & \kt\\
&       4       &7764.7791 $\pm$        0.0040 &        2 $\pm$ 14 & \kt\\
&       5       &7770.1865 $\pm$        0.0156 &        5  $\pm$ 26 & \kt\\
&       6       &7775.5908 $\pm$        0.0067 &        4 $\pm$ 16 & \kt\\
&       7       &7780.9950 $\pm$        0.0063 &        2 $\pm$ 16 & \kt\\
&       8       &7786.3989 $\pm$        0.0098 &        0 $\pm$ 19 & \kt\\
&       9       &7791.8063 $\pm$        0.0104 &        3 $\pm$ 20 & \kt\\
&       10      &7797.1958 $\pm$        0.0175 &        $-20$ $\pm$ 28 & \kt\\
&       11      &7802.6102 $\pm$        0.0080 &        $-7$ $\pm$ 17 & \kt\\
&       12      &7808.0162 $\pm$        0.0088 &        $-6$ $\pm$ 18 & \kt\\
&       13      &7813.3996 $\pm$        0.0304 &  $-38$ $\pm$ 46 & \kt\\
&       182     &8726.8223 $\pm$        0.0081 &        $-158$ $\pm$ 14 & B22\\
&       194     &8791.7629 $\pm$        0.0066 &        $-48$ $\pm$ 12 & B22\\
&       249     &9089.0881 $\pm$        0.0062 &        $-4\pm$ 12 & \cheops\\
&       252     &9105.3154 $\pm$        0.0032 &        12 $\pm$ 10 & \cheops\\
&       254     &9116.1251 $\pm$        0.0031 &        10 $\pm$ 10 & \cheops\\
&       256     &9126.9414 $\pm$        0.0067 &        18 $\pm$ 13 & \cheops\\
\hline
\end{tabular}
\tablefoot{Sources of reference epochs (T$_0$) and transit mid-times correspond to  \citet{christiansen2018} (Ch18), \citet{lopez2019} (L19), \citet{hardegree-ullman2021} (HU21) and \citet{boyle2022} (B22).}
\end{table*}

\begin{table*}[]
 \caption{Transit times and deviations of the $e$, $f$, and $g$ planets in the $K$2-138 system.}
 \label{tab:ttvs_continuacion}
  \centering
    \begin{tabular}{ c c c c l}
        \hline
        \hline
  Planet & Epoch & T$_{\rm C}-$2450000 &O$-$C & Source \\
               &       & [BJD\_TDB] & [min] & \\
        \hline
$e$ &0 &        7740.6449  $\pm$        0.0028  & $-1$ $\pm$ 4 & \kt\\
&       1 &     7748.9067  $\pm$        0.0032  & $-1$ $\pm$ 5 &  \kt \\
&       2 &     7757.1691 $\pm$ 0.0027  & 1 $\pm$ 4 & \kt \\
&       3 &     7765.4292 $\pm$ 0.0032  & $-1$ $\pm$ 5 & \kt \\
&       4 &     7773.6804 $\pm$ 0.0241  & $-16$ $\pm$ 35 & \kt \\
&       5 &     7781.9516 $\pm$ 0.0050  & $-1$ $\pm$ 7 & \kt \\
&       7 &     7798.4757 $\pm$ 0.0040  & 1 $\pm$ 6 & \kt \\
&       8 &     7806.7375 $\pm$ 0.0027  & 1 $\pm$ 4 & \kt \\
&       9 &     7815.0029 $\pm$ 0.0076  & 7 $\pm$ 11 & \kt \\
&       162 &   9078.9850 $\pm$ 0.0035  & 4 $\pm$ 6 & \cheops\\
&       163 &   9087.2384 $\pm$ 0.0082  & $-7$ $\pm$ 12 & \cheops\\
&       167 &   9120.2834 $\pm$ 0.0055  & $-8$ $\pm$ 8 & \cheops\\
\hline
$f$ &   0 &     7738.7032 $\pm$ 0.0089  & $-3$ $\pm$ 15 & \kt\\
&       1 &     7751.4591 $\pm$ 0.0076  & $-4$ $\pm$ 13 &  \kt \\
&       2 &     7764.2184 $\pm$ 0.0043  & 0 $\pm$ 10 &  \kt \\
&       3 &     7776.9743 $\pm$ 0.0061  & $-1$ $\pm$ 11 &  \kt \\
&       5 &     7802.4941 $\pm$ 0.0069  & 8 $\pm$ 12 &  \kt \\
&       6 &     7815.2451 $\pm$ 0.0128  & 0 $\pm$ 20 &  \kt \\
&       103 &   9052.6488 $\pm$ 0.0041  & 17 $\pm$ 13 & \cheops \\
&       104 &   9065.3823 $\pm$ 0.0097  & $-16$ $\pm$ 18 & \cheops\\
&       105 &   9078.1351 $\pm$ 0.0035  & $-22$ $\pm$ 13 & \cheops\\
&       106 &   9090.9090 $\pm$ 0.0075  & 3 $\pm$ 16 & \cheops\\
&       108 &   9116.4282 $\pm$ 0.0036  & 12 $\pm$ 13 & \cheops\\
&       109 &   9129.1660 $\pm$ 0.0172  & $-15$ $\pm$ 28 & \cheops\\
\hline
$g$ & 0 &       7773.8541       $\pm$ 0.0044 &  $-2$ $\pm$ 7 & \kt\\
&       1  &    7815.8205       $\pm$ 0.0044 &  0 $\pm$ 7 &  \kt \\
&       10 &    8193.5155       $\pm$ 0.0084 &  8 $\pm$ 12  & HU21\\
&       32 &    9116.7492       $\pm$ 0.0020 &  0 $\pm$ 5 & \cheops\\
\hline
\end{tabular}
\end{table*}

\begin{figure}[h]
    \centering
    \includegraphics[width=\linewidth]{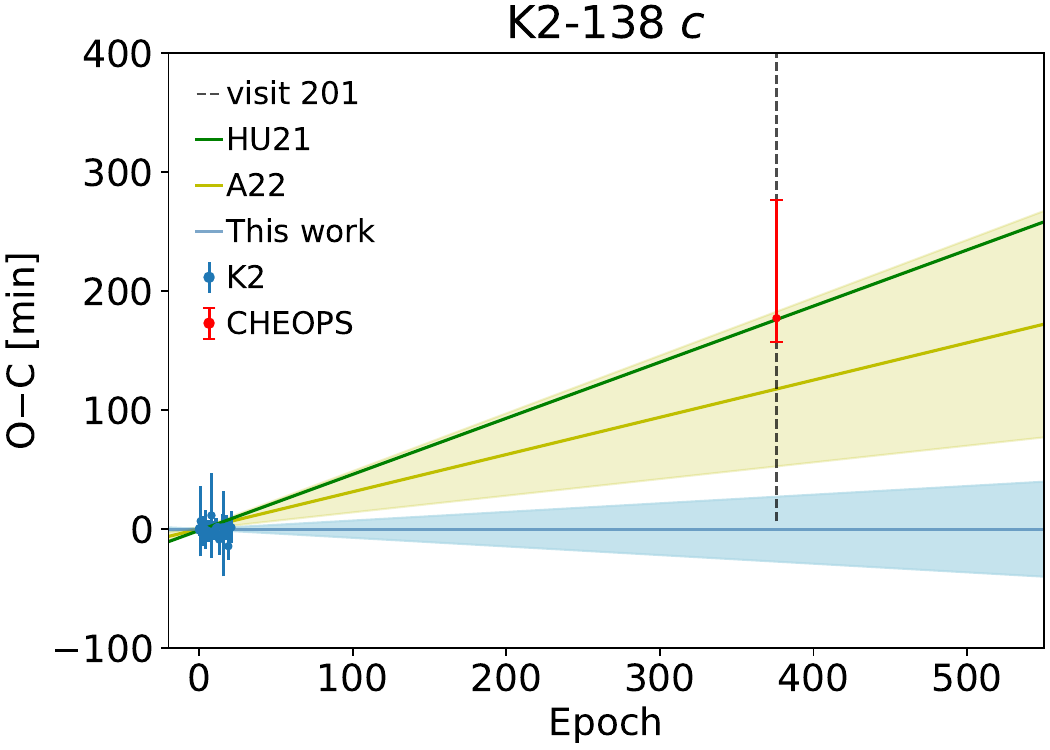}
    \caption{Observed minus calculated (O-C) diagram of the transit times of planet $c$. The estimates from \kt{} (blue symbols) are compared to our updated ephemeris, which uncertainties are represented by the light blue region. The predicted times of the ephemeris from HU21 (green curve) and A22 (yellow curve and region) are also shown. The full duration of the \cheops{} visit 201 is marked with the vertical dashed curve and the corresponding fitted mid-time assuming a transit detection (see Sect.~\ref{ssec:planet_c}) is marked with the red symbol.}
    \label{fig:OC_c}
\end{figure}

\begin{figure}[h]
    \centering
    \includegraphics[width=\linewidth]{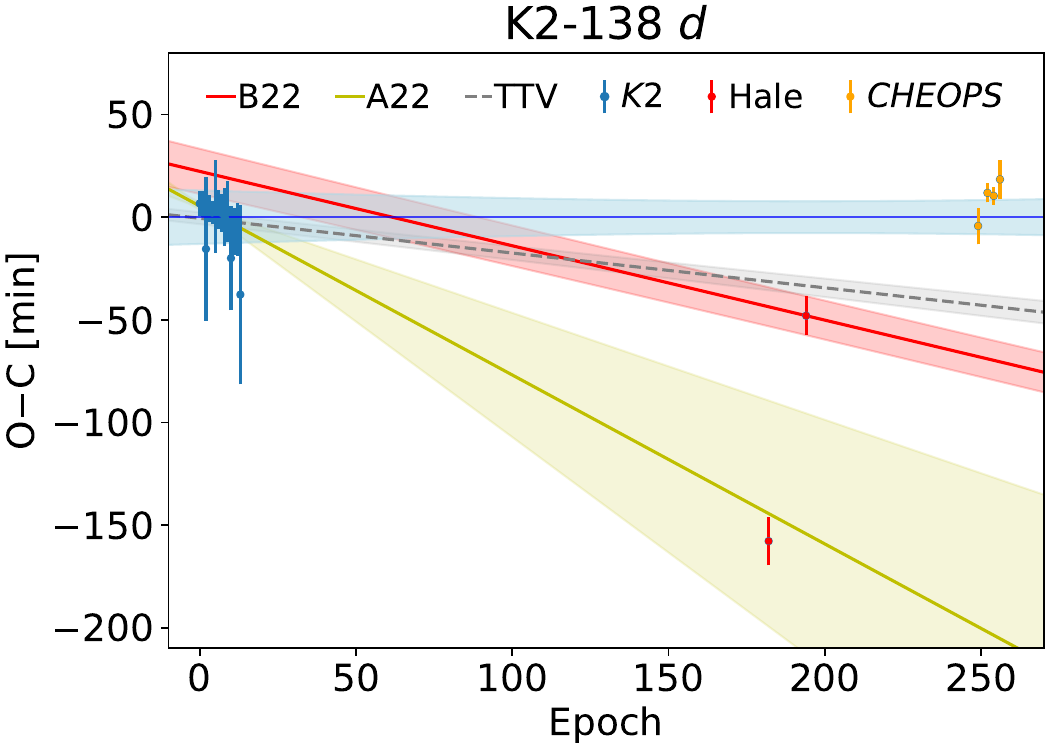}
    \caption{Observed minus calculated (O-C) diagram of the transit times of K2-328\,$d$, compared to our updated ephemeris (light blue region), as described in Sect.~\ref{ssec:planet_d}. The times of the transits observed by \kt{}, Hale observatory, and \cheops{} are shown with the blue, red, and orange symbols, respectively. The predicted ephemeris from A22 and B22 are represented with yellow and red curves and regions. The mean ephemeris obtained by the TTV simulations (Sect.~\ref{ssec:ttv_modeling}) is represented by the gray dashed curve.}
    \label{fig:OC_d_v2}
\end{figure}

\begin{figure}[h]
    \centering
    \includegraphics[width=\linewidth]{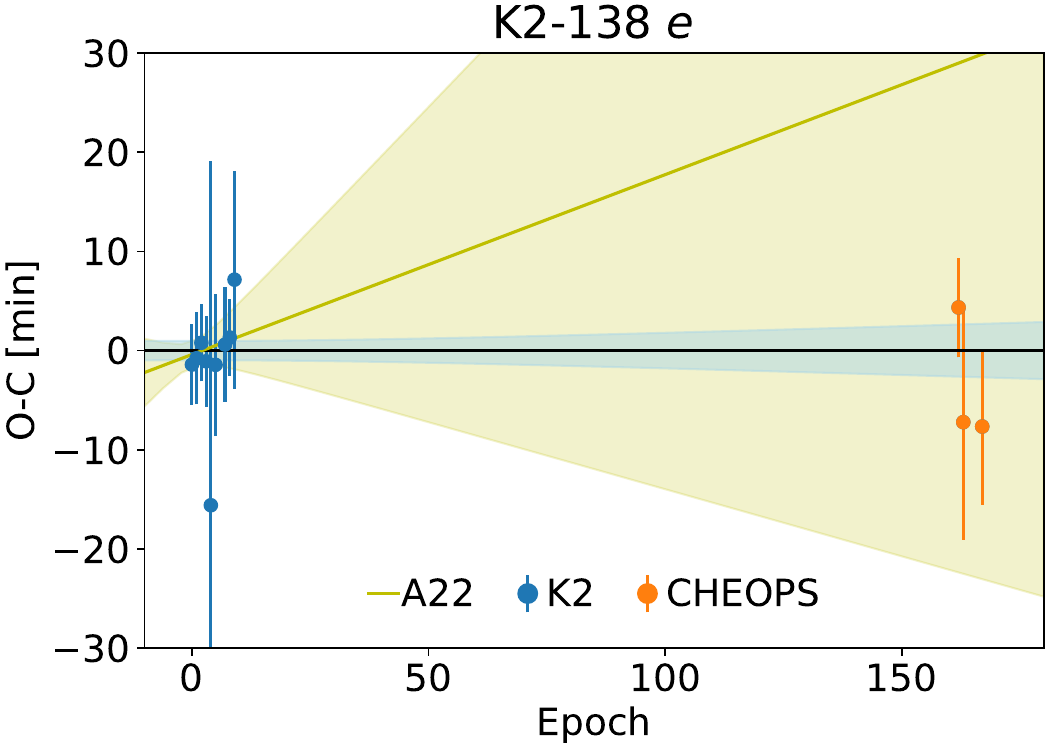}
    \caption{Observed minus calculated (O-C) diagram of the transit times of planet $e$. The \kt{} and \cheops{} transit times are shown with the blue and orange symbols and compared with our updated ephemeris (light blue region). The yellow curve and region represent the ephemeris of A22 and their uncertainties. }
    \label{fig:OC_e}
\end{figure}

\begin{figure}[h]
    \centering
    \includegraphics[width=\linewidth]{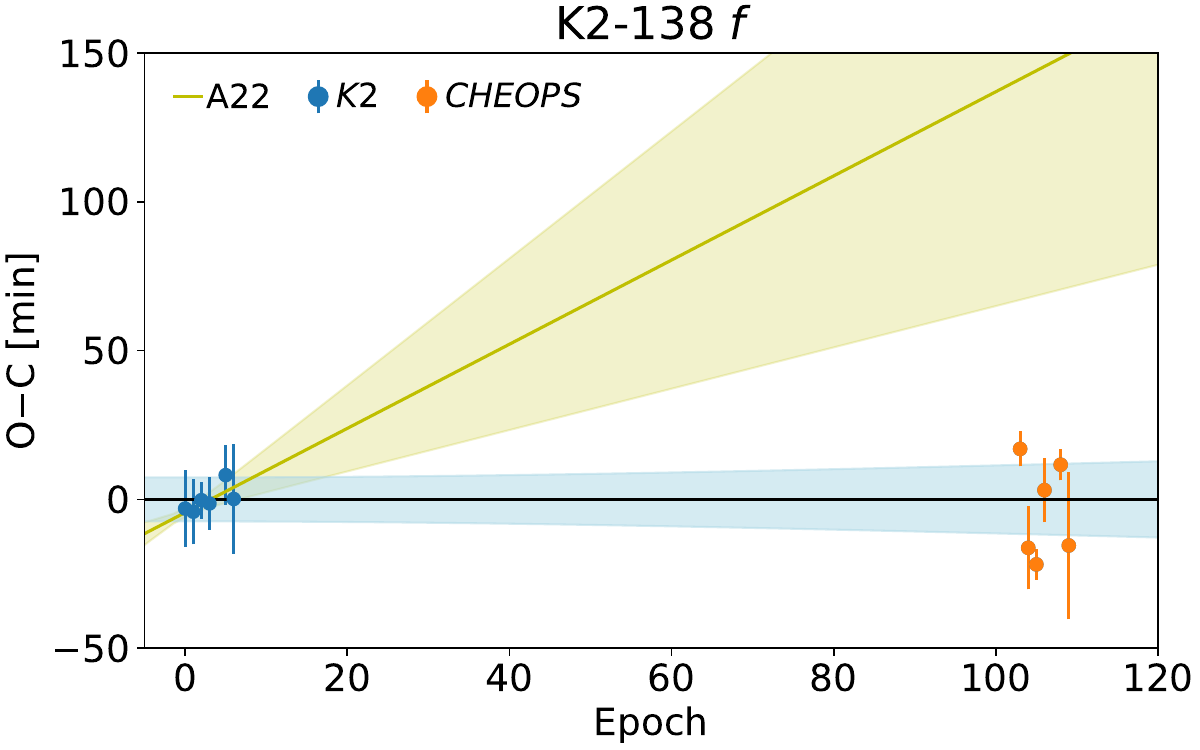}
    \caption{Observed minus calculated (O-C) diagram of the transit times of planet $f$. The \kt{} and \cheops{} transit times (blue and orange symbols) were used to refine the ephemeris reported by A22 (yellow curve and region). The uncertainties of the updated ephemeris are represented by the light blue region. }
    \label{fig:OC_f}
\end{figure}

\begin{figure}[h]
    \centering
    \includegraphics[width=\linewidth]{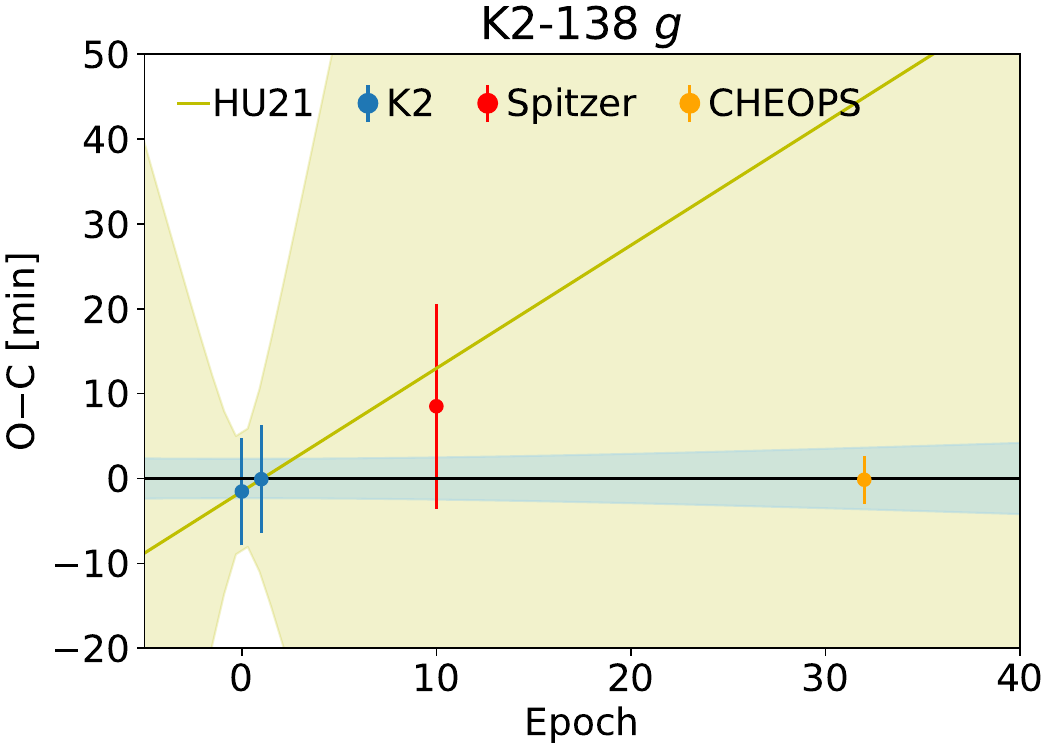}
    \caption{Observed minus calculated (O-C) diagram of the transit times of planet $g$. The \kt{}, \spitzer{} and \cheops{} transit times (blue, red and orange symbols) allowed to refine the ephemeris reported by HU21, yellow curve and region). The uncertainties of our updated ephemeris are represented by the light blue region. }
    \label{fig:OC_g}
\end{figure}

\end{appendix}
\end{document}